\documentclass[aps,12pt,superscriptaddress,amsfonts,amssymb,amsmath]{revtex4}
\pdfoutput=1

\usepackage{graphicx}
\usepackage{epsfig}
\usepackage{makeidx}
\usepackage{epstopdf}
\usepackage{natbib}
\usepackage{xcolor}
\usepackage{subcaption}
\usepackage{amsmath}
\usepackage{appendix}

\begin{document}

\title{Gauge waves generation and detection in Aharonov-Bohm electrodynamics}
\author{F.\ Minotti \footnote{Email address: minotti@df.uba.ar}}
\affiliation{Universidad de Buenos Aires, Facultad de Ciencias Exactas y Naturales, Departamento de F\'{\i}sica, Buenos Aires, Argentina}
\affiliation{CONICET-Universidad de Buenos Aires, Instituto de F\'{\i}sica Interdisciplinaria y Aplicada (INFINA), Buenos Aires, Argentina}

\author{G.\ Modanese \footnote{Email address: giovanni.modanese@unibz.it}}
\affiliation{Free University of Bozen-Bolzano \\ Faculty of Engineering \\ I-39100 Bolzano, Italy}

\linespread{0.9}

\begin{abstract}
The extended Aharonov-Bohm electrodynamics has a simple formal structure and allows to couple the e.m.\ field also to currents which are not locally conserved, like those resulting from certain non-local effective quantum models of condensed matter. As it often happens in physics and mathematics when one tries to extend the validity of some equations or operations, new perspectives emerge in comparison to Maxwell theory, and some ``exotic'' phenomena are predicted. For the Aharonov-Bohm theory the main new feature is that the potentials $A^\mu$ become univocally defined and can be measured with probes in which the ``extra-current'' $I=\partial_\mu j^\mu$ is not zero at some points. As a consequence, it is possible in principle to detect pure gauge-waves with $\mathbf{E}=\mathbf{B}=0$, which would be regarded as non-physical in the Maxwell gauge-invariant theory with local current conservation. We discuss in detail the theoretical aspects of this phenomenon and propose an experimental realization of the detectors. A full treatment of wave propagation in anomalous media with extra-currents and of energy-momentum balance issues is also given.
\end{abstract}

\maketitle

\section{Introduction}

Although classical and quantum electrodynamics are among the most precise and successful theories in physics, there are reasons to look for extensions of Maxwell equations in various directions and physical contexts \cite{bandos2020nonlinear,sorokin2022introductory}. In particular, we are interested into the extension first proposed by Ohmura \cite{ohmura1956new} and often called ``Aharonov-Bohm electrodynamics'' because of their major contribution (\cite{aharonov1963further}; see also \cite{van2001generalisation,hively2012toward,Modanese2017MPLB,modanese2017electromagnetic,arbab2017extended,hively2019classical,Hively_2021,minotti2021quantum,minotti2022electromagnetic}). The logical motivation for this extension, at least from today's perspective, is the quest for a Lagrangian density which has the most general covariant form, in terms of four-potentials $A_\mu$ and that sources $j^\mu$ (not necessarily locally conserved). Apart from the addition of a total four-divergence, this form is
\begin{equation}
 L_{A.-B.}=\frac{1}{4\mu_0} F_{\mu\nu}F^{\mu\nu}-\frac{\lambda}{2\mu_0} (\partial_\mu A^\mu)^2+A_\mu j^\mu
 \label{LAB}
 \end{equation}
 where a possible additional photon mass term proportional to $A_\mu A^\mu$ has been excluded based on experimental evidence.

 \subsection{Formal structure of local Aharonov-Bohm electrodynamics}
 
 We shall call (\ref{LAB}) the ``Aharonov-Bohm Lagrangian'', even though it is in fact a well-known expression contained in textbooks (see e.g.\ \cite{itzykson2012quantum,fradkin2021quantum}) and its unicity has been proven by Woodside \cite{woodside2009three} in vector formalism. This Lagrangian is gauge-invariant in the familiar sense only if $\lambda=0$. Therefore by considering extensions with non-zero $\lambda$ we are  sacrificing gauge invariance to extended relativistic invariance and especially to the possibility of coupling $A_\mu$ to a current $j^\mu$ which is not locally conserved. 
 
 Maxwell theory with full gauge invariance can be of course re-obtained from this extended theory in the special case when $\lambda=0$ and current is locally conserved everywhere. The power of gauge theories in quantum field theory and particle physics of course strongly suggests to make this  symmetrical choice whenever possible.

 The special choice $\lambda=1$ is known in the literature as ``Feynman gauge'' (although it is not a gauge condition in the usual sense) and leads to a quantization method which is alternative to the Gupta-Bleuler method or to canonical quantization \cite{fradkin2021quantum}. As we show in the following, in the extended Aharonov-Bohm theory this choice is natural and almost mandatory in the presence of currents which are not locally conserved.
 
 The field equations derived from the Lagrangian (\ref{LAB}) are
 \begin{equation}
  \partial^2 A^\mu + (\lambda-1) \partial^\mu\partial_\nu A^\nu=\mu_0 j^\mu
 \label{LABeq1}
 \end{equation}
 with $\partial^2=\partial_\mu \partial^\mu$. From this one has by contraction
 \begin{equation}
  \partial^2 (\partial_\nu A^\nu)= \frac{\mu_0}{\lambda}I
 \label{LABeq2}
 \end{equation}
 where $I$ is called ``extra-current'' and defined as
 \begin{equation}
  I=\partial_\mu j^\mu
 \label{defI}
 \end{equation}
 This allows to rewrite eq.\ (\ref{LABeq1}), for $\lambda \neq 0$, as
 \begin{equation}
  \partial^2 A^\mu= \mu_0 j^\mu + \mu_0 \frac{1-\lambda}{\lambda} \partial^\mu [(\partial^2)^{-1} I]
 \label{LABeq3}
 \end{equation}
 
 Now, on physical grounds it is very tempting to add the condition that in these equations the fields are ``locally forced'' by the physical sources $j^\mu$, in the sense that second derivatives of the fields at one point in spacetime are determined by the values of the sources at that point. All fundamental theories for massless fields give, at least in their linearized versions, wave equations for the field which have this property.

 In this case, this is accomplished by setting $\lambda=1$. As a consequence, the four-potentials of Aharonov-Bohm theory resemble those of Maxwell theory in Lorentz gauge or Feynman gauge, and in fact we shall express them routinely as retarded integrals over the sources. There is a clear difference, however, apparent from eq.\ (\ref{LABeq2}): if charge is not locally conserved everywhere, it is not possible in general to impose the gauge condition $\partial_\nu A^\nu=0$. Actually, there is no gauge freedom in the AB theory, except for the possibility of reduced gauge transformations of the form $A^\mu \to A^\mu+\partial^\mu \chi$, where the function $\chi$ satisfies $\partial^2 \chi=0$.

 Therefore in most physical situations where the physical sources are localized, the potentials are completely determined. As we shall see in the following, they can even be measured using hypothetical material probes in which $\partial_\mu j^\mu \neq 0$ ! All other ``standard'' probes can only measure the fields $F^{\mu\nu}$ (i.e., $\mathbf{E}$ and $\mathbf{B}$).

 From the discussion above it is also clear that in A.-B. electrodynamics there is a sort of additional degree of freedom, namely a scalar field $S=\partial_\mu A^\mu$ satisfying the equation $\partial^2 S=\mu_0 I$. It is not really an independent degree of freedom because it is expressed in terms of $A_\mu$, but it appears explicitly in the extended field equations. The scalar field becomes automatically decoupled from the sources if $I = 0$.

 We do not think that $S$ can have any relevant physical effects in normal laboratory conditions, because we hypothesize that violations of local charge conservation can only occur in microscopic quantum sources whose emission is very weak at the macroscopic scale (the situation may be different at an astrophysical or cosmological scale, see e.g.\ \cite{jimenez2011cosmological}). Nevertheless, precise measurements could allow to detect the generation of an $S$ field \cite{modanese2019design,minotti2021current}. Also, in order to check the general consistency of the theory, we have computed the propagation of $S$ in material media and found that it is completely reflected by ohmic conductors and pure dielectrics (see \cite{minotti2023aharonov}, preferably in the arXiv version).

\begin{table}
\begin{center}
\textbf{Local AB electrodynamics at a glance} \\
\begin{tabular}{|c|c|} 
\hline
Fundamental fields & $A^\mu$ \nonumber \\
Derived fields & $F^{\mu\nu}=\partial^\mu A^\nu-\partial^\nu A^\mu $; \ $S=\partial_\mu A^\mu$ \nonumber \\
Lagrangian & $\frac{1}{4\mu_0}F^{\mu\nu}F_{\mu\nu}-\frac{1}{2\mu_0} S^2+A_\mu j^\mu$ \nonumber \\
Field equations & $\partial^2 A^\mu= \mu_0 j^\mu$ \nonumber \\
Equation for $S$ & $\partial^2 S= \mu_0 I$, \ $I=\partial_\mu j^\mu$ \nonumber \\
Reduced gauge invariance & $A^\mu \to A^\mu+\partial^\mu \chi$, \ $\partial^2 \chi=0$ \nonumber \\
Lorenz force 4-density on probe $p$ & $f^\mu=F^\mu_\nu j_p^\nu-I_pA^\mu$ \nonumber \\
Vector Lorenz force and power & $\mathbf{f}=\rho_p\mathbf{E}+\mathbf{j}_p\times \mathbf{B}-I_p\mathbf{A}$, \ $w=\mathbf{j}_p\cdot \mathbf{E}-I_p\phi$ \nonumber \\
\hline
\end{tabular}	
\caption{Formal framework of Aharonov-Bohm extended electrodynamics with the choice $\lambda=1$ which ensures local field equations. The ``minimalistic'' structure of the theory appears: given the currents $j^\mu$, one directly solves for $A^\mu$ and computes the Lorenz four-force density on a probe with current $j^\mu_p$ and extra-current $I_p$. The four-force density is given here also in vector form in view of the simple application to a $\phi$-probe described later; the relation between four- and three-vectors is $A^\mu=(\phi/c,\mathbf{A})$, $j^\mu=(c\rho,\mathbf{j})$. For the field equations for $F^{\mu\nu}$ and a comparison with Maxwell theory in the Lorentz gauge see Sect.\ \ref{comp}.
}		
\label{table1}
\end{center}
\end{table}

\subsection{Energy conservation. Lorenz force}
\label{ene}

The conceptual structure of the theory is summarized in Tab.\ \ref{table1} and will be further discussed in Sect.\ \ref{comp} in comparison with Maxwell theory. In order to complete the picture we still need the conservation laws of energy and momentum of the extended e.m.\ field and the generalization of the formula for the Lorenz force. 

The conservation laws have been derived in [MM-2021] by coupling the field to a generic gravitational background. The calculation is complex, but the result is relatively simple and the energy density $u$, for example, can be written in vector formalism as
\begin{equation}
    u=\frac{1}{\mu_0} \left( \frac{\mathbf{E}^2}{2c^2} + \frac{\mathbf{B}^2}{2} + \frac{\phi}{c^2} \frac{\partial S}{\partial t} - \mathbf{A} \cdot \nabla S - \frac{S^2}{2} \right)
\end{equation}
Some contributions depending on $S$ and explicitly dependent on the potentials can be seen here. 

For the Lorenz force density and for the power exerted on a probe $p$ with charge density $\rho_p$, current density $\mathbf{j}_p$ and extra-current $I_p$ one finds
\begin{eqnarray}
    \mathbf{f}&=&\rho_p\mathbf{E}+\mathbf{j}_p\times \mathbf{B}-I_p\mathbf{A} \\
    w&=&\mathbf{j}_p\cdot \mathbf{E}-I_p\phi
\end{eqnarray}
This means that anomalous extra-sources not only generate a (usually small) scalar field contributing to the energy density, but can also feel the effect of the potentials $\mathbf{A}$, $\phi$, and act as probes for measuring them. It is thus confirmed that the Aharonov-Bohm potentials should be regarded as physical. In Sect.\ \ref{possible} we will study a possible concrete realization of a $\phi$-detector.

\subsection{Field equations. Comparison with Maxwell theory}
\label{comp}

Although the conceptual structure of local Aharonov-Bohm electrodynamics described so far and summarized in Tab.\ \ref{table1} is consistent and very simple, it is also useful to look at the equations for the field strengths $F^{\mu\nu}$ and to compare the theory with the version of Maxwell theory obtained by imposing the Lorentz gauge condition $\partial_\mu A^\mu=0$.

The equations for $F^{\mu\nu}$ derived from the Lagrangian (\ref{LAB}) have the following form \cite{Modanese2017MPLB}, remarkably independent from $\lambda$:
\begin{equation}
    \partial_\mu F^{\mu\nu} =\mu_0 (j^\nu+i^\nu)
    \label{eqF}
\end{equation}
where $i^\nu$ is an additional four-current defined by $i^\nu=\partial^\nu \partial^{-2} I=\partial^\nu S/\mu_0$. The antisymmetry of $F^{\mu\nu}$ implies the total conservation $\partial_\mu(j^\mu+i^\mu)=0$. (This is reminiscent of what happens with the chiral anomaly in Weyl systems, where the anomalous term can be rewritten as a current that satisfies the classical balance equation \cite{morawetz2019chiral,morawetz2019weyl}.)
In Sect.\ \ref{gamma} the equations for $F^{\mu\nu}$ are written explicitly in vector form, in order to apply them to materials with violation of local conservation in the so-called ``$I$-$\gamma$-$j$'' model.

In spite of the reduced gauge invariance of the Aharonov-Bohm Lagrangian, since the homogeneous field equations $\nabla \times \mathbf{E}=-\partial_t \mathbf{B}/c$ and $\nabla \cdot \mathbf{B}=0$ are the same as in Maxwell theory, it is possible to write $\mathbf{E}$ and $\mathbf{B}$ in terms of a non-physical, auxiliary four-potential, whose gauge condition can be freely chosen, as usual in Maxwell theory. This property is quite useful, since it allows to employ most traditional techniques also for the extended theory.

At first sight the  statement above might appear to lead to a contradiction with our previous finding that in the extended theory it is impossible in general to impose the Lorentz gauge. Moreover, since in Maxwell theory the choice of the Lorentz gauge allows to write the equations for the potentials just in the desired locally forced form, one might be tempted to conclude that the local Aharonov-Bohm theory is nothing but Maxwell theory in Lorentz gauge plus the (apparently arbitrary) assumption that the four-potential is physical. 

This is actually not true, because an auxiliary four-potential in Lorentz gauge for the field equations (\ref{eqF}) in fact exists but satisfies the non-local equations
\begin{equation}
    \partial^2 A^\mu_{aux,Lor}=\mu_0 (j^\mu-\partial^\mu \partial^{-2}I)
\end{equation}
and not the equation for the physical four-potential $A^\mu$, which is $\partial^2 A^\mu=\mu_0 j^\mu$ ! Therefore there are no contradictions. Instead, we can see here two strong connections between Maxwell theory and the extended AB theory: 

(1) In applications where the extra-current $I$ can be disregarded as source and considered only as probe, the  auxiliary potentials $A^\mu_{aux,Lor}$ and the physical potentials $A^\mu$ essentially coincide (see Sect.\ \ref{defgw}). 

(2) If one is only interested into the fields $\mathbf{E}$ and $\mathbf{B}$, they can be computed using any gauge condition for the auxiliary potentials.

\subsection{In which physical systems can the extra-current $I$ be non-zero?}
\label{phys}

The physical systems where a violation of local conservation might occur, yielding $I\neq 0$ in some regions, are thought to be mainly the following:

\begin{enumerate}

\item Complex condensed-matter systems described by a quantum field theory, in which the local conservation of the current operator is spoiled by anomalies occurring in the renormalization process \cite{cheng1984gauge,parameswaran2014probing}.

\item Molecular devices, like e.g.\ carbon nanotubes and other molecular ``wires'', in which the effect of bound electrons in the inner orbitals upon the conduction electrons is well modelled through a non-local potential, and the anomaly is not due to the use of a reduced eigenstates base, but remains at any order in the computations \cite{walz2015local,jensen2019current,garner2019helical,garner2020three}.

\item Systems with explicitly non-local wave equations, e.g.\ fractional quantum mechanics and other phenomenological models \cite{lenzi2008solutions, lenzi2008fractional, latora1999superdiffusion, caspi2000enhanced, chamon1997nonlocal, balantekin1998green, laskin2002fractional, wei2016comment,modanese2018time}.

\item Macroscopic quantum systems like Josephson junctions and superconductors in general. For a system with macroscopic wavefunction, the charge density and current density are represented by non-commuting operators. It follows that the extra-current $I=\partial_t \rho + \nabla \cdot \mathbf{j}$, being essentially a linear combination of these two operators in the frequency-momentum domain, does not admit eigenstates and has a minimum uncertainty fixed by the Heisenberg relation $\Delta N \Delta \phi \simeq 1$ which involves the occupation number and the phase of the wavefunction. In \cite{minotti2021quantum} we have given an estimate of the minimum uncertainty in the case of a tunnel Josephson junction made of Nb.

\end{enumerate}

 \subsection{Definition and properties of gauge waves}
\label{defgw}

We define a gauge wave as a plane wave with potentials
\begin{eqnarray}
    \phi&=&\phi_0 e^{i(\omega t - \mathbf{k}\mathbf{x})} \label{p1a}\\
    \mathbf{A}&=& \mathbf{A}_0 e^{i(\omega t - \mathbf{k}\mathbf{x})}
    \label{p1b}
\end{eqnarray}
for which the relation 
\begin{equation}
    \omega \mathbf{A}_0=\mathbf{k}\phi_0
    \label{p2}
\end{equation}
holds. It is therefore a longitudinal wave, with Fourier amplitudes $\phi_0$ and $\mathbf{A}_0$ connected in absolute value by $|\phi_0|=c|\mathbf{A}_0|$, as follows from (\ref{p2}) and from $\omega/|\mathbf{k}|=c$. It is a solution of the AB equations in vacuum $\partial^2 A^\mu=0$.

Taking the scalar product by $\mathbf{k}$ in eq.\ (\ref{p2}) we also find that
\begin{equation}
    \mathbf{k} \cdot \mathbf{A}_0=\frac{\omega}{c^2}\phi_0
    \label{p3}
\end{equation}
This means that the potentials defined by (\ref{p1a}), (\ref{p1b}) satisfy the equation
\begin{equation}
    \nabla \cdot \mathbf{A}-\frac{1}{c^2} \frac{\partial \phi}{\partial t}=0
    \label{p4}
\end{equation}
which is the vector equivalent of the Lorentz gauge condition $\partial_\mu A^\mu=0$.

Remembering the definition of the scalar field $S$ as $S=\partial_\mu A^\mu$ we thus see that $S=0$ in a gauge wave. It is also straightforward to prove that in this wave $\mathbf{E}$ and $\mathbf{B}$ are zero, using the general relations between fields and potentials, namely
\begin{eqnarray}
    \mathbf{E}&=&\nabla \phi -\frac{1}{c} \frac{\partial \mathbf{A}}{\partial t}=i(\mathbf{k}\phi_0-\omega \mathbf{A}_0)e^{i(\omega t - \mathbf{k}\mathbf{x})}=0 \\
    \mathbf{B}&=&\nabla\times\mathbf{A}=i(\mathbf{k}\times\mathbf{A}_0)e^{i(\omega t - \mathbf{k}\mathbf{x})}=0
    \label{p5}
\end{eqnarray}
This property justifies the name chosen for the gauge wave.

Since the fields $S$, $\mathbf{E}$ and $\mathbf{B}$ are all zero in a gauge wave, the wave does not carry any energy and momentum, according to the general expressions in \cite{minotti2021quantum} (for the energy flux see also eq.\ (\ref{energy_flux})). In a gauge-invariant theory it would be regarded as non-physical, also because it would be equivalent, via a gauge transformation, to a four-potential which is identically zero. We know, however, that in AB electrodynamics the potentials are physical and interact with an anomalous matter probe with extra-current $I_p$ through the terms $I_p\phi$ and $I_p\mathbf{A}$ in the Lorenz force density.

In \cite{minotti2023aharonov} we have proven that, consistently with the picture above, a gauge wave propagates without attenuation in normal media where local charge conservation holds. In Sect.\ ... of this work the propagation of general plane waves in anomalous media is analysed using the ``$\gamma$-model'' approximation for the anomalous current density.

\subsection{Generation of gauge waves}
\label{generation_gw}

Concerning the generation of gauge waves, in \cite{minotti2023aharonov} it was shown that when a plane wave of the $S$ field encounters a normal medium, it cannot enter it but is totally reflected, possibly generating a gauge wave at the interface. This process, however, is not the most efficient way of producing gauge waves, since it requires an incident $S$ wave, which can only be generated by an anomalous oscillating source (see computation in Sect.\ \ref{radiating}).

It is thus far more convenient to take advantage of the gauge wave components which are generated in normal oscillating dipoles and go unnoticed unless they impinge on a material probe containing an extra-current $I_p$. 

In fact, the physical AB potentials of an oscillating dipole are (like in Maxwell theory in Lorentz gauge)
\begin{eqnarray}
    \phi(\mathbf{x},t) &=& \frac{\mu_0 c}{4\pi r} \dot{\mathbf{p}}(t-r/c)\cdot \mathbf{n} \\
    \mathbf{A}(\mathbf{x},t) &=& \frac{\mu_0}{4\pi r} \dot{\mathbf{p}}(t-r/c)
    \label{dip1}
\end{eqnarray}
where $r=|\mathbf{x}|$ and $\mathbf{n}=\mathbf{x}/r$ is the local propagation direction, such that the spherical wave can be approximated at each point by a plane wave with vector $\mathbf{k}=|\mathbf{k}|\mathbf{n}$.

The vector $\mathbf{A}$ can be decomposed into a longitudinal component
\begin{equation}
    \mathbf{A}_\mathbf{n}=(\mathbf{A}\cdot\mathbf{n})\mathbf{n}=\frac{\phi}{c} \mathbf{n}
\end{equation}
satisfying the gauge wave condition $|\mathbf{A}_\mathbf{n}|=\phi/c$, and a transverse component $\mathbf{A}_T$. 

Along the oscillation axis of the dipole we have that $\mathbf{A}_T=0$, so that the wave is longitudinal and a purely gauge wave. At any other point, the four-potential can be considered as the sum of a gauge wave $(\phi,\mathbf{A}_\mathbf{n})$ and a transverse vector potential $(0,\mathbf{A}_T)$. If the gauge wave encounters a normal conducting medium it is not affected (supposing the process is linear). The transverse component, on the other hand, generally interacts with the medium and in several cases can be strongly dampened, leaving as a result a pure gauge wave which propagates in the medium and then possibly gets out of the medium again into free space.

A hypothetical detection antenna for gauge waves will be described in Sect.\ \ref{possible}.

\subsection{Outline}
After this long Introduction, the rest of the work will be organized as follows. Sect.\ \ref{gamma} is devoted to what we call the ``$\gamma$-model'', namely a simplified one-parameter model for the description of material media containing currents non locally conserved. The non-dimensional parameter $\gamma$ quantifies in a well-defined sense the average, macroscopic effect of the conservation violations which occur at the microscopic level. In Sect.\ \ref{free-c} the model is applied to free charges, in Sect.\ \ref{diel} to dielectric media, where $\gamma$ appears in the relations among $\mathbf{j}$, $\mathbf{P}$ (local polarization vector) and $\mathbf{M}$ (local magnetization vector). As a consequence, in Sect.\ \ref{macr-AB} we write the extended AB field equations in a medium and in Sect.\ \ref{wave-p} the wave propagation equations in the medium. Sect.\ \ref{possible} contains a discussion of possible observable effects of AB extended electrodynamics. In Sect.\ \ref{detection} we describe a basic DC circuit for detection of gauge-waves. It contains a voltage generator and a resistor made of a material with conservation violation parameter $\gamma$. The scalar potential $\phi$ of a gauge-wave can interact with the fraction $\gamma$ of the resistor current that is not locally conserved. Using one of the fundamental relations of the $\gamma$-model, namely $\langle \phi I \rangle = \gamma \langle \mathbf{j}\cdot \nabla\phi \rangle +$ surface term, and imposing energy conservation and conditions for the surface term to be negligible, one can prove that an additional voltage $\delta V$ arises in the resistor, of order $\delta V \sim \gamma \Delta \phi$, where $\Delta \phi$ is the variation of the potential $\phi$ across the conductor length. Sect.\ \ref{antenna} contains a description of the possible measurable effects of gauge waves. We first describe the basic scheme of a DC circuit for the detection of a generic potential gradient $\Delta \phi$ through an extra-current flowing in a conductor. Then we consider an antenna detector and its response to an harmonic gauge wave propagating along the direction of the antenna. 
Sect.\ \ref{radiating} contains a general calculation of radiation fields generated by harmonic currents in anomalous conductors with negligible polarization described by the $\gamma$ model. Fields are computed in the dipole approximation. An evaluation of the time-averaged energy flow at large distance from the source closes the section. Finally, Sect.\ \ref{concl} contains our Conclusions.

\section{Material media with non-conserved currents: the $\protect\gamma $
model}
\label{gamma}

In order to build up a theory of Aharonov-Bohm (AB) extended electrodynamics
in material media with non-conserved currents we need to reconsider the
usual models for the macroscopic description of fields and sources in those
media.

To begin, we start with the AB equations in vacuum 
\begin{subequations}
\label{ABeqsvaccum}
\begin{eqnarray}
\nabla \cdot \mathbf{E} &=&\frac{\rho }{\varepsilon _{0}}-\frac{\partial S}{ 
\partial t},  \label{ABGauss} \\
\nabla \cdot \mathbf{B} &=&0,  \label{ABmonopole} \\
\nabla \times \mathbf{E} &=&-\frac{\partial \mathbf{B}}{\partial t},
\label{ABFaraday} \\
\nabla \times \mathbf{B} &=&\mu _{0}\varepsilon _{0}\frac{\partial \mathbf{E} 
}{\partial t}+\mu _{0}\mathbf{j}+\nabla S,  \label{ABAmpere} \\
\mathbf{E} &=&-\nabla \phi -\frac{\partial \mathbf{A}}{\partial t},
\label{ABEpots} \\
\mathbf{B} &=&\nabla \times \mathbf{A},  \label{ABBpots} \\
S &=&\mu _{0}\varepsilon _{0}\frac{\partial \phi }{\partial t}+\nabla \cdot 
\mathbf{A},  \label{ABScalar}
\end{eqnarray}
\end{subequations}

From Eqs. (\ref{ABeqsvaccum}) it results that the extra source $I$, defined as 
\begin{equation*}
I=\frac{\partial \rho }{\partial t}+\nabla \cdot \mathbf{j},
\end{equation*}
is the source of the scalar, which satisfies 
\begin{equation}
\mu _{0}\varepsilon _{0}\frac{\partial ^{2}S}{\partial t^{2}}-\nabla
^{2}S=\mu _{0}I.  \label{wavescalar}
\end{equation}

These equations are valid if all sources $\varrho $ and $\mathbf{j}$ are
known, which is the main difficulty in dealing with material media. The
usual approach is then to consider for all magnitudes macroscopic averages
(introduced below), that should average out the small scale, rapid
fluctuations of the sources and fields. Furthermore, the sources are
separated into free and molecular. Molecular sources are those due to
microscopic charges and currents that are highly localized in space, in or
around the molecules they belong to. Free charges, on the other hand, are
those that are not bounded to a particular molecule and can more or less
freely move in the medium.

\subsection{Free charges}
\label{free-c}

In the model presented in \cite{minotti2022electromagnetic}, which we refer here as the $I$-$\gamma$-$j$ model, or simply $\gamma$ model, it is
considered that a likely candidate to account for non-local conservation of
charge in material media is the tunnel effect of electrons. If a microscopic
free electron current of value $i_{n}$ discontinuously crosses from the
point $\mathbf{x}_{n}$ to the point $\mathbf{x}_{n}+\mathbf{d}_{n}$, the
corresponding extra source is 
\begin{equation}
I_{n}=i_{n}\left[ \delta \left( \mathbf{x}-\mathbf{x}_{n}-\mathbf{d} 
_{n}\right) -\delta \left( \mathbf{x}-\mathbf{x}_{n}\right) \right] .
\label{Inbase}
\end{equation} 
This relation results from the definition of the extra source $I$ and the
semiclassical conceptualization of instantaneous tunneling as the
disappearance of the current at $\mathbf{x}_{n}$ with its simultaneous
reappearance at $\mathbf{x}_{n}+\mathbf{d}_{n}$, which corresponds to a
spatial redistribution of charge without an accompanying electric current. It is to be noted that extra sources with similar opposite peaks appear in the solution of the Schr\"odinger equation with non-local interactions \cite{modanese2018time}.

Furthermore, in the $I$-$\gamma$-$j$ model the tunneled current $i_{n}$ in the
volume element $d^{3}x=dSdL$ represents a fraction of the current $\mathbf{j} 
\cdot d\mathbf{S}$ traversing it, while the tunneled distance is a fraction
of the volume length $dL$, so that $i_{n}\mathbf{d}_{n}=\gamma \mathbf{j}d^{3}x$, with $\gamma$ the product of those fractions. In a sense, the $I$-$\gamma$-$j$ model represents the simplest phenomenological model one can conceive based on a single parameter.

The idea is now to consider a macroscopically small volume $\delta \Omega $
centered at position $\mathbf{x}=\mathbf{X}$, with dimensions much larger
than the characteristic tunneling distance $\left\vert \mathbf{d}
_{n}\right\vert $, which is taken as a measure of the microscopic scale. For
a generic microscopic field $F\left( \mathbf{x},t\right) $ one thus defines
a macroscopic corresponding field by spatial averaging (at each time) of the
form 
\begin{equation}
\left\langle F\right\rangle _{\mathbf{X}}=\frac{1}{\delta \Omega }
\int_{\delta \Omega }F\left( \mathbf{x},t\right) d^{3}x,  \label{genavg}
\end{equation}
where $\delta \Omega $ is the same for all $\mathbf{X}$.

Immediate properties from the very definition (\ref{genavg}) are 
\begin{eqnarray*}
\frac{\partial }{\partial t}\left\langle F\right\rangle _{\mathbf{X}} &=& 
\frac{1}{\delta \Omega }\int_{\delta \Omega }\frac{\partial }{\partial t}
F\left( \mathbf{x},t\right) d^{3}x, \\
\frac{\partial }{\partial X_{i}}\left\langle F\right\rangle _{\mathbf{X}}
&=& \frac{1}{\delta \Omega }\int_{\delta \Omega }\frac{\partial }{\partial
x_{i}} F\left( \mathbf{x},t\right) d^{3}x.
\end{eqnarray*}

Using these properties, since Eqs. (\ref{ABeqsvaccum}) are linear their
average results in identical equations, but now applied to the corresponding
macroscopic magnitudes.

Special care must be taken when considering the macroscopic current density,
because, according to the $\gamma$ model, part of charge is redistributed in
a microscopically discontinuous manner. Indeed, if one considers the
electric charge contained in a given elementary volume 
\begin{equation*}
\delta Q=\int_{\delta \Omega }\rho \left( \mathbf{x},t\right) d^{3}x,
\end{equation*}
its time variation depends on the net flow of charge across $\Sigma \left(
\delta \Omega \right) $, the boundary of $\delta \Omega $, 
\begin{equation*}
\frac{\partial }{\partial t}\delta Q=-\oint_{\Sigma \left( \delta \Omega
\right) }\mathbf{j}\left( \mathbf{x},t\right) \cdot d\mathbf{S}
+\sum_{n}i_{n},
\end{equation*}
where $\mathbf{j}\left( \mathbf{x},t\right) $ represents the flow of charge
that is transported continuously, and $\sum_{n}i_{n}$ represents the sum of
the elementary currents that tunnel across $\Sigma \left( \delta \Omega
\right) $ (with their corresponding sign, according to whether they enter or
leave the volume).

Using Gauss theorem and the average properties we can write 
\begin{eqnarray*}
\frac{1}{\delta \Omega }\oint_{\Sigma \left( \delta \Omega \right) }\mathbf{ 
j }\left( \mathbf{x},t\right) \cdot d\mathbf{S} &=&\frac{1}{\delta \Omega }
\int_{\delta \Omega }\nabla _{\mathbf{x}}\cdot \mathbf{j}\left( \mathbf{x}
,t\right) d^{3}x \\
&=&\nabla _{\mathbf{X}}\cdot \left\langle \mathbf{j}\right\rangle _{\mathbf{ 
X }},
\end{eqnarray*}
so that the macroscopic charge density $\left\langle \rho \right\rangle _{ 
\mathbf{X}}=\delta Q/\delta \Omega $ satisfies the relation 
\begin{equation}
\frac{\partial }{\partial t}\left\langle \rho \right\rangle +\nabla _{ 
\mathbf{X}}\cdot \left\langle \mathbf{j}\right\rangle =\frac{1}{\delta
\Omega }\sum_{n}i_{n}.  \label{drhodt1}
\end{equation}

In order to proceed we expand on the $I-\gamma-j $ model. In this model the tunneled current $i_{n}$ in a given volume element is a fraction of the normal current flowing through it, and the total tunneled distance is also a fraction of the length traversed by the normal current in that volume. Furthermore, we take into consideration that in order that a given $i_{n}$ contributes to the charge transport across $\Sigma \left( \delta \Omega \right) $ the corresponding tunneled length $\mathbf{d}_{n}$ must cross the surface (with either $\mathbf{x}_{n}$ or $\mathbf{x}_{n}+\mathbf{d}_{n}$ outside $\delta\Omega $). In this way, if normal current crosses a given surface element ($\mathbf{j}\cdot d\mathbf{S}\neq 0$), the probability that tunneled currents cross the surface element is given by the fraction of tunneled to normal current, times the fraction of tunneled distance to total distance (the latter accounts for the probability that $\mathbf{d}_{n}$ crosses the surface). Both mentioned fractions make up the coefficient $\gamma $, so that the current that actually tunnels across a given surface element is $\gamma\mathbf{j}\cdot d\mathbf{S}$. We thus have
\begin{eqnarray}
\frac{1}{\delta \Omega }\sum_{n}i_{n} &=&-\frac{1}{\delta \Omega } 
\oint_{\Sigma \left( \delta \Omega \right) }\gamma \mathbf{j}\left( \mathbf{x 
},t\right) \cdot d\mathbf{S}  \notag \\
&=&-\frac{1}{\delta \Omega }\int_{\delta \Omega }\gamma \nabla _{\mathbf{x} 
}\cdot \mathbf{j}\left( \mathbf{x},t\right) d^{3}x=-\gamma \nabla _{\mathbf{X 
}}\cdot \left\langle \mathbf{j}\right\rangle .  \label{sumin}
\end{eqnarray}

Using (\ref{drhodt1}) and (\ref{sumin}), the relation between macroscopic free
charge and current densities is finally expressed as 
\begin{equation}
\frac{\partial }{\partial t}\left\langle \rho \right\rangle +\left( 1+\gamma
\right) \nabla _{\mathbf{X}}\cdot \left\langle \mathbf{j}\right\rangle =0.
\label{drhodtfinal}
\end{equation}

A related important magnitude is the macroscopic extra source 
\begin{equation*}
\left\langle I\right\rangle _{\mathbf{X}}=\frac{1}{\delta \Omega }
\int_{\delta \Omega }\sum_{n\subseteq \delta \Omega }I_{n}d^{3}x,
\end{equation*}
which is determined, according to (\ref{Inbase}), by only those microscopic
currents for which the corresponding tunneled length $\mathbf{d}_{n}$
crosses the surface with either $\mathbf{x}_{n}$ or $\mathbf{x}_{n}+\mathbf{ 
d }_{n}$ outside $\delta \Omega $. This gives precisely the same result as (\ref{sumin}) 
\begin{equation}
\left\langle I\right\rangle _{\mathbf{X}}=-\gamma \nabla _{\mathbf{X}}\cdot
\left\langle \mathbf{j}\right\rangle .  \label{Imacro}
\end{equation}

An important result of the $I-\gamma-j $ model, from (\ref{drhodtfinal}) and (\ref{Imacro}), is that for a stationary case the macroscopic extra source
due to free charges is zero: $\left\langle I\right\rangle _{\mathbf{X}}=0$ if $\partial \left\langle \rho \right\rangle / \partial t=0$

Finally, it will also be necessary to consider averages of the type 
\begin{equation*}
\left\langle FI\right\rangle _{\mathbf{X}}=\frac{1}{\delta \Omega } 
\int_{\delta \Omega }F\left( \mathbf{x}\right) \left( \sum_{n\subseteq
\delta \Omega }I_{n}\right) d^{3}x,
\end{equation*} 
for different fields $F\left( \mathbf{x}\right) $. Using (\ref{Inbase}) we have 
\begin{eqnarray*}
\left\langle FI\right\rangle _{\mathbf{X}} &=&\frac{1}{\delta \Omega } 
\sum_{n\subseteq \delta \Omega }i_{n}\left[ F\left( \mathbf{x}_{n}+\mathbf{d} 
_{n}\right) -F\left( \mathbf{x}_{n}\right) \right]  \\
&&+\frac{1}{\delta \Omega }\sum_{n\subseteq \Sigma \left( \delta \Omega
\right) }sg_{n}i_{n}F\left( \mathbf{x}_{n}\right) .
\end{eqnarray*} 
where the second term in the right-hand side accounts for the contribution
of those microscopic currents for which the corresponding tunneled length $ 
\mathbf{d}_{n}$ crosses the surface with either $\mathbf{x}_{n}$ or $\mathbf{ 
x}_{n}+\mathbf{d}_{n}$ outside $\delta \Omega $, and $sg_{n}$ is the
corresponding sign, positive when $i_{n}$ enters the volume, and negative
when it leaves it. We now write 
\begin{equation*}
F\left( \mathbf{x}_{n}+\mathbf{d}_{n}\right) -F\left( \mathbf{x}_{n}\right)
=\nabla F\cdot \mathbf{d}_{n},
\end{equation*} 
and use the $I-\gamma-j $ model as before, which allows us to write the
contributions of the volume and of the surface terms as 
\begin{equation}
\left\langle FI\right\rangle _{\mathbf{X}}=\frac{1}{\delta \Omega } 
\int_{\delta \Omega }\gamma \mathbf{j}\cdot \nabla Fd^{3}x-\frac{1}{\delta
\Omega }\oint_{\Sigma \left( \delta \Omega \right) }\gamma F\mathbf{j}\cdot d 
\mathbf{S}. \label{FIXgeneral}
\end{equation} 
Using Gauss theorem in the surface term we finally obtain 
\begin{equation}
\left\langle FI\right\rangle _{\mathbf{X}}=-\frac{1}{\delta \Omega } 
\int_{\delta \Omega }\gamma F\nabla \cdot \mathbf{j}d^{3}x=-\gamma
\left\langle F\nabla \cdot \mathbf{j}\right\rangle _{\mathbf{X}}.
\end{equation}

Care must be taken in the evaluation of averages of this form due to their
being non-linear in the microscopic fields. To proceed we follow the
formalism in \cite{Minotti-PRE-2000} in which the averages of the form (\ref 
{genavg}) are complemented with the definition of the fluctuations, relative
to the averages, as 
\begin{equation*}
\delta F\left( \mathbf{x},\mathbf{X},t\right) \equiv F\left( \mathbf{x} 
,t\right) -\left\langle F\right\rangle _{\mathbf{X}}.
\end{equation*}

This particular definition of fluctuations satisfies the relation 
\begin{equation}
\left\langle \delta F\left( \mathbf{x},\mathbf{X},t\right) \left\langle
G\right\rangle _{\mathbf{X}}\right\rangle _{\mathbf{X}}=0,  \label{avgfluct}
\end{equation} 
for any two fields $F\left( \mathbf{x},t\right) $ and $G\left( \mathbf{x} 
,t\right) $. In this way, by writing 
\begin{eqnarray*}
F\left( \mathbf{x},t\right) &=&\left\langle F\right\rangle _{\mathbf{X} 
}+\delta F\left( \mathbf{x},\mathbf{X},t\right) , \\
\nabla \cdot \mathbf{j}\left( \mathbf{x},t\right) &=&\nabla _{\mathbf{X} 
}\cdot \left\langle \mathbf{j}\right\rangle _{\mathbf{X}}+\delta \nabla
\cdot \mathbf{j}\left( \mathbf{x},\mathbf{X},t\right)
\end{eqnarray*} 
we have 
\begin{equation*}
\left\langle F\nabla \cdot \mathbf{j}\right\rangle _{\mathbf{X} 
}=\left\langle F\right\rangle _{\mathbf{X}}\nabla _{\mathbf{X}}\cdot
\left\langle \mathbf{j}\right\rangle _{\mathbf{X}}+\left\langle \delta
F\delta \nabla \cdot \mathbf{j}\right\rangle _{\mathbf{X}}.
\end{equation*}

The average of fluctuations in this expression results from two
contributions: that due to fluctuations of scale much smaller than the
macroscopic one, $\delta L$, and that due to fluctuations of scale close to $ 
\delta L$. The latter contribution can be determined using the method in 
\cite{Minotti-PRE-2000}, while the contributions of fluctuations of very
small scale need to be modeled independently. The fluctuations of scale not
too small contribute to a generic average of fluctuations $\left\langle
\delta F\delta G\right\rangle $ a term \cite{Minotti-PRE-2000} 
\begin{equation*}
\left\langle \delta F\delta G\right\rangle \simeq \frac{\left( \delta
L\right) ^{2}}{24}\nabla _{\mathbf{X}}\left\langle F\right\rangle \cdot
\nabla _{\mathbf{X}}\left\langle G\right\rangle ,
\end{equation*} 
and since $\left\langle F\right\rangle $ and $\left\langle G\right\rangle $\
are macroscopic fields their variation over $\delta L$ is small, 
\begin{equation*}
\left\vert \frac{\left( \delta L\right) ^{2}}{24}\nabla _{\mathbf{X} 
}\left\langle F\right\rangle \cdot \nabla _{\mathbf{X}}\left\langle
G\right\rangle \right\vert \ll \left\vert \left\langle F\right\rangle
\left\langle G\right\rangle \right\vert ,
\end{equation*} 
so that one can safely neglect this contribution. Concerning the
contribution of the scales much smaller than $\delta L$, we will also
neglect it considering that the level of those fluctuations is not supposed
to be highly excited above their thermal level, in the absence of a
mechanisms leading to a non-linear energy cascade. In this way, we finally
obtain 
\begin{equation}
\left\langle FI\right\rangle _{\mathbf{X}}=-\gamma \left\langle
F\right\rangle _{\mathbf{X}}\nabla _{\mathbf{X}}\cdot \left\langle \mathbf{j} 
\right\rangle _{\mathbf{X}}.  \label{FIavgfinal}
\end{equation}

We can now analyze the application of Ohm law. The point is that in \cite 
{minotti2021quantum} the following expression for the force density on matter due to the fields was obtained 
\begin{subequations}
\label{interacfieldmatter}
\begin{equation*}
\mathbf{f}=\rho \mathbf{E}+\mathbf{j}\times \mathbf{B}-I\mathbf{A}.
\end{equation*}
\end{subequations}

Using the previous considerations we obtain 
\begin{eqnarray}
\left\langle I\mathbf{A}\right\rangle _{\mathbf{X}} &=&-\gamma \left\langle 
\mathbf{A}\right\rangle _{\mathbf{X}}\nabla _{\mathbf{X}}\cdot \left\langle 
\mathbf{j}\right\rangle _{\mathbf{X}},  \label{IAavg} \\
\left\langle \mathbf{j}\times \mathbf{B}\right\rangle _{\mathbf{X}}
&=&\left\langle \mathbf{j}\right\rangle_{\mathbf{X}} \times \left\langle \mathbf{B} 
\right\rangle_{\mathbf{X}} =\left\langle \mathbf{j}\right\rangle_{\mathbf{X}} \times \left( \nabla _{ 
\mathbf{X}}\times \left\langle \mathbf{A}\right\rangle_{\mathbf{X}} \right) .
\label{jBavg}
\end{eqnarray} 
This allows us to estimate the relative order of magnitude of these two
terms, considering that the averaged magnitudes have spatial variations of the same relative order. This shows that the additional force coming from the non-conserved
currents is a term of small magnitude (for $\gamma $ small) compared to the
Lorentz force. In this way, we take as a reasonable approximation the usual
relation between local electric field and total current to be $\left\langle 
\mathbf{j}\right\rangle_{\mathbf{X}} =\sigma \mathbf{E}$, with $\sigma $ the conductivity
of the medium, even if it has to be extended to include magnetic forces, like in the Hall effect.

\subsection{Dielectric media}
\label{diel}

We consider now the macroscopic relations for polarizable media, in which
the molecular sources play a fundamental role.

To begin, the usual expression of averaged molecular charge density (no
explicit distinction between the spatial coordinates of volume centers and
space coordinates would be made from now on) 
\begin{equation}
\left\langle \rho _{m}\right\rangle =-\nabla \cdot \mathbf{P},  \label{rhom}
\end{equation}
expressed in terms of a vector $\mathbf{P}$, is justified considering that
the total molecular charge in a medium without added free charges should be
zero, irrespective of the shape of the medium. Besides, the interpretation
of $\mathbf{P}$ as the volume density of electric polarization results from
expression (\ref{rhom}) and the definition of the dipole moment of the
medium as $\int \left\langle \rho _{m}\right\rangle \mathbf{x}d^{3}x$, with
the integral extended to the whole medium.

We thus see that expression (\ref{rhom}) can be taken over for the
description of media with non-conserved currents. Furthermore, the
phenomenological relation $\mathbf{P}=\varepsilon _{0}\varkappa \left\langle 
\mathbf{E}\right\rangle $ for a linear, isotropic medium, with a
polarization coefficient $\varkappa $, should also be an acceptable model in
our case, since it does not conflict with any assumption in the $\gamma$
model.

We now consider the macroscopic averages $\mathbf{J}_{m}$ of molecular
currents. The condition that molecular currents are highly localized implies
that their flux across any macroscopic cross section of the medium should be
zero, which translates into the relation $\mathbf{J}_{m}=$ $\nabla \times 
\mathbf{M}$ for a generic vector $\mathbf{M}$. This is valid under the
assumption that the spatial averaging process cancels the rapidly
fluctuating in time, instantaneous unbalances of molecular currents crossing
the surface element. However, when the macroscopic fields are not
stationary, a time variation of a polarization vector $\mathbf{P}$\ with
non-zero divergence corresponds to an orderly redistribution of charges,
crossing the boundary of the element, but remaining localized in the
molecules that cross that boundary. This effect gives rise to a contribution
to the macroscopic molecular current of the form $\partial \mathbf{P}
/\partial t$. It is thus customary to include both contributions in the
expression 
\begin{equation*}
\mathbf{J}_{m}=\frac{\partial \mathbf{P}}{\partial t}+\nabla \times \mathbf{ 
M }.
\end{equation*}

Moreover, for stationary situations, the interpretation of $\mathbf{M}$ as
the volume density of magnetic moment results from the very definition of
total magnetic moment of the medium: $\frac{1}{2}\int \mathbf{x}\times 
\mathbf{J}_{m}d^{3}x$, with the integral extended to the whole medium, and
the expression of $\mathbf{J}_{m}$ valid in this case.

Furthermore, for linear, isotropic media the idea of the orientation of the
molecular magnetic dipoles along the magnetic field leads to a model with a
linear relation between $\mathbf{M}$ and $\mathbf{B}$, usually expressed
through the auxiliary vector $\mathbf{H}=\left( \left\langle \mathbf{B}
\right\rangle -\mathbf{M}\right) /\mu _{0}$ as $\mathbf{M}=\chi \mathbf{H}$,
where $\chi $ is the magnetic susceptibility coefficient. This latter
phenomenological model can also be taken over to our case, since it is fully
compatible with the $\gamma$ model.

Note that the macroscopic molecular sources so far considered satisfy the
continuity relation 
\begin{equation*}
\frac{\partial }{\partial t}\left\langle \rho _{m}\right\rangle +\nabla
\cdot \mathbf{J}_{m}=0.
\end{equation*}
However, we must take into account that in the spirit of the $\gamma$ model
some of the molecular charge redistribution can take place discontinuously,
even within the molecule in which it is localized. The likely candidate for
this is the redistribution of charge leading to variation of the dipole
moment in each molecule. For this reason, we assume that only a fraction $ 
\alpha $ of $\partial \mathbf{P}/\partial t$ is due to a continuous charge
motion. We thus consider that an effective fraction $\gamma ^{\prime }$\ (in
general different from the $\gamma $ for free currents) of the continuous
current crosses a given surface element discontinuously, so that 
\begin{equation*}
\left\langle I\right\rangle _{\mathbf{X}}=-\gamma ^{\prime }\alpha \nabla
\cdot \frac{\partial \mathbf{P}}{\partial t}.
\end{equation*}

These considerations lead to a redefinition of the continuous component $ 
\mathbf{J}_{m}$ as 
\begin{equation}
\mathbf{J}_{m}=\alpha \frac{\partial \mathbf{P}}{\partial t}+\nabla \times 
\mathbf{M},  \label{Jm}
\end{equation}
so that now the continuity relation is 
\begin{equation*}
\frac{\partial }{\partial t}\left\langle \rho _{m}\right\rangle +\nabla
\cdot \mathbf{J}_{m}=-\gamma ^{\prime }\alpha \nabla \cdot \frac{\partial 
\mathbf{P}}{\partial t},
\end{equation*}
which using also (\ref{rhom}) is written as 
\begin{equation*}
-\frac{\partial }{\partial t}\nabla \cdot \mathbf{P}+\alpha \nabla \cdot 
\frac{\partial \mathbf{P}}{\partial t}=-\gamma ^{\prime }\alpha \nabla \cdot 
\frac{\partial \mathbf{P}}{\partial t},
\end{equation*}
which requires that 
\begin{equation*}
\alpha =\frac{1}{1+\gamma ^{\prime }}.
\end{equation*}
resulting in the macroscopic extra source contributed by the polarization of
the medium to be 
\begin{equation}
\left\langle I\right\rangle _{\mathbf{X}}=-\frac{\gamma ^{\prime }}{1+\gamma
^{\prime }}\nabla \cdot \frac{\partial \mathbf{P}}{\partial t}.
\label{ImacroPol}
\end{equation}

As for the contribution from free charges to the macroscopic extra source,
also that due to molecular sources turns out to be zero in stationary
situations.

\subsection{Macroscopic AB equations in media with non-conserved currents}
\label{macr-AB}

We now recall the relations which, according to the considerations in the
previous sections, could serve as a valid model for the macroscopic fields
and sources in a linear, homogeneous medium with non-locally conserved
currents.

We first recall that, due to their being linear, the volume average
procedure can be applied directly to Eqs. (\ref{ABeqsvaccum}) to yield
exactly the same equations, but where all fields and sources are to be
considered as macroscopic ones. Besides, the sources are to be considered as
including both components, free and molecular.

Furthermore, for easy of notation, from now on all macroscopic fields and sources are represented without the average symbol $\left\langle...\right 
\rangle $. The spatial coordinate $\mathbf{x}$ is also not distinguished from the center of the volume elements, previously designated as $\mathbf{X}$.

With these considerations, the molecular sources are given by 
\begin{eqnarray*}
\rho _{m} &=&-\nabla \cdot \mathbf{P}, \\
\mathbf{j}_{m} &=&\frac{1}{1+\gamma ^{\prime }}\frac{\partial \mathbf{P}}{
\partial t}+\nabla \times \mathbf{M},
\end{eqnarray*}
where the polarization and magnetization have the usual expressions 
\begin{eqnarray*}
\mathbf{P} &=&\left( \varepsilon -\varepsilon _{0}\right) \mathbf{E}, \\
\mathbf{M} &=&\left( 1/\mu _{0}-1/\mu \right) \mathbf{B}.
\end{eqnarray*}

Using the previous relations to express the macroscopic molecular sources we
obtain the equations 
\begin{subequations}
\label{ABmedia}
\begin{eqnarray}
\nabla \cdot \mathbf{E} &=&\frac{\rho }{\varepsilon }-\frac{\varepsilon _{0} 
}{\varepsilon }\frac{\partial S}{\partial t},  \label{gauss} \\
\nabla \cdot \mathbf{B} &=&0,\;\mathbf{B}=\nabla \times \mathbf{A},
\label{nomonop} \\
\nabla \times \mathbf{E} &=&-\frac{\partial \mathbf{B}}{\partial t},\; 
\mathbf{E}=-\nabla \phi -\frac{\partial \mathbf{A}}{\partial t},
\label{faraday} \\
\nabla \times \mathbf{B} &=&\mu \mathbf{j}+\mu \varepsilon \frac{1+\gamma
^{\prime }\frac{\varepsilon _{0}}{\varepsilon }}{1+\gamma ^{\prime }}\frac{
\partial \mathbf{E}}{\partial t}+\frac{\mu }{\mu _{0}}\nabla S,
\label{ampere} \\
S &=&\mu _{0}\varepsilon _{0}\frac{\partial \phi }{\partial t}+\nabla \cdot 
\mathbf{A}.  \label{scalar}
\end{eqnarray}
\end{subequations}

The sources appearing explicitly in these relations are only the free
sources, the molecular ones having been included through their expressions
in terms of the fields.

The equations for the potentials can be more simply derived from the
original system (\ref{ABeqsvaccum}), which after further averaging and use
of the expressions of the molecular sources in terms of the fields gives 
\begin{subequations}
\label{potentials}
\begin{eqnarray}
\mu _{0}\varepsilon _{0}\frac{\varepsilon _{0}}{\varepsilon }\frac{\partial
^{2}\phi }{\partial t^{2}}-\nabla ^{2}\phi &=&\frac{\rho }{\varepsilon }
+\left( 1-\frac{\varepsilon _{0}}{\varepsilon }\right) \frac{\partial }{
\partial t}\left( \nabla \cdot \mathbf{A}\right) ,  \label{potscal} \\
\mu \varepsilon \frac{1+\gamma ^{\prime }\frac{\varepsilon _{0}}{\varepsilon 
}}{1+\gamma ^{\prime }}\frac{\partial ^{2}\mathbf{A}}{\partial t^{2}}-\nabla
^{2}\mathbf{A} &=&\mu \mathbf{j}+\left( \frac{\mu }{\mu _{0}}-1\right)
\nabla \left( \nabla \cdot \mathbf{A}\right)  \notag \\
&&+\mu \varepsilon \frac{\frac{\varepsilon _{0}}{\varepsilon }-1}{1+\gamma
^{\prime }}\nabla \left( \frac{\partial \phi }{\partial t}\right) .
\label{potvect}
\end{eqnarray}
\end{subequations}
The previous equations must be complemented with the relation for free
sources, Eq. (\ref{drhodtfinal}), 
\begin{equation*}
\frac{\partial \rho }{\partial t}+\left( 1+\gamma \right) \nabla \cdot 
\mathbf{j}=0,
\end{equation*}
where the free current is related to the electric field by the Ohm law 
\begin{equation}
\mathbf{j}=\sigma \mathbf{E}=-\sigma \left( \nabla \phi +\frac{\partial 
\mathbf{A}}{\partial t}\right) .  \label{OhmLaw}
\end{equation}

\subsection{Wave propagation}
\label{wave-p}

We study in this section the possible plane-wave modes in which all fields
and sources are of the generic form $a_{0}\exp \left[ i\left( \mathbf{k}
\cdot \mathbf{x}-\omega t\right) \right] $, with constant complex amplitude $ 
a_{0}$.

From Eqs. (\ref{drhodtfinal}) and (\ref{OhmLaw}) we have for the mode
amplitudes of the free sources 
\begin{eqnarray*}
\varrho _{0} &=&-i\frac{\sigma \left( 1+\gamma \right) }{\omega }\left(
k^{2}\phi _{0}-\omega \mathbf{k}\cdot \mathbf{A}_{0}\right) , \\
\mathbf{j}_{0} &=&-i\sigma \left( \mathbf{k}\phi _{0}-\omega \mathbf{A}
_{0}\right) ,
\end{eqnarray*}
which are used in Eqs. (\ref{potentials}) to finally obtain the relations
for the potentials amplitudes 
\begin{subequations}
\label{amplitudepots}
\begin{eqnarray}
\left\{ k^{2}\left[ 1+i\frac{\sigma \left( 1+\gamma \right) }{\omega
\varepsilon }\right] -\mu _{0}\varepsilon _{0}\frac{\varepsilon _{0}}{
\varepsilon }\omega ^{2}\right\} \phi _{0} &=&\omega \left[ 1-\frac{
\varepsilon _{0}}{\varepsilon }+i\frac{\sigma \left( 1+\gamma \right) }{
\omega \varepsilon }\right] \mathbf{k}\cdot \mathbf{A}_{0}, \\
\left( k^{2}-\mu \varepsilon \frac{1+\gamma ^{\prime }\frac{\varepsilon _{0} 
}{\varepsilon }}{1+\gamma ^{\prime }}\omega ^{2}-i\mu \sigma \omega \right) 
\mathbf{A}_{0} &=&\left( 1-\frac{\mu }{\mu _{0}}\right) \mathbf{k}\left( 
\mathbf{k}\cdot \mathbf{A}_{0}\right)  \notag \\
&&+\left( \mu \varepsilon \frac{\frac{\varepsilon _{0}}{\varepsilon }-1}{
1+\gamma ^{\prime }}\omega -i\mu \sigma \right) \mathbf{k}\phi _{0}.
\end{eqnarray}
\end{subequations}

The transverse mode ($\mathbf{k}\cdot \mathbf{A}_{0}=0$) corresponds to a
wave with $\phi _{0}=0$ and dispersion relation 
\begin{equation*}
k^{2}=\mu \varepsilon \frac{1+\gamma ^{\prime }\frac{\varepsilon _{0}}{
\varepsilon }}{1+\gamma ^{\prime }}\omega ^{2}+i\mu \sigma \omega .
\end{equation*}

For a dielectric without losses ($\sigma =0$) this is an electromagnetic
wave with phase speed given by 
\begin{equation*}
c=\sqrt{\mu \varepsilon \frac{1+\gamma ^{\prime }\frac{\varepsilon _{0}}{
\varepsilon }}{1+\gamma ^{\prime }}},
\end{equation*}
which indicates a slight (for $\gamma ^{\prime }\ll 1$) decrease of the
usual speed of light in the medium.

For the case of a conducting medium there is also the usual decay associated
to Joule dissipation, which is very strong in good conductors with $\frac{
\sigma }{\omega \varepsilon }\gg 1$.

For the longitudinal modes ($\mathbf{k}\times \mathbf{A}_{0}=0$) we obtain
the dispersion relation

\begin{equation*}
\left( \frac{k^{2}}{\mu _{0}\varepsilon _{0}\omega ^{2}}-1\right) \left\{
\left( \frac{k^{2}}{\mu _{0}\varepsilon _{0}\omega ^{2}}-1\right) \left[ 1+i 
\frac{\sigma \left( 1+\gamma \right) }{\omega \varepsilon }\right] +\frac{
\gamma ^{\prime }\left( 1-\frac{\varepsilon _{0}}{\varepsilon }\right) }{
1+\gamma ^{\prime }}+i\frac{\gamma \sigma }{\omega \varepsilon }\right\} =0.
\end{equation*}

We thus see that the branch $k^{2}=\mu _{0}\varepsilon _{0}\omega ^{2}$,
corresponding to a wave of pure potentials, or gauge wave, which has zero
fields ($\mathbf{E}$, $\mathbf{B}$ and $S$), is allowed for all values of $ 
\gamma $, $\gamma ^{\prime }$ and $\sigma $.

The other branch has the dispersion relation 
\begin{equation}
\frac{k^{2}}{\mu _{0}\varepsilon _{0}\omega ^{2}}=1-\frac{\gamma ^{\prime
}\left( 1-\frac{\varepsilon _{0}}{\varepsilon }\right) +i\frac{\gamma \sigma 
}{\omega \varepsilon }\left( 1+\gamma ^{\prime }\right) }{\left( 1+\gamma
^{\prime }\right) \left[ 1+i\frac{\sigma \left( 1+\gamma \right) }{\omega
\varepsilon }\right] }.  \label{disprelS}
\end{equation}

When relation (\ref{disprelS}) is used in expressions (\ref{amplitudepots}),
the relation between the amplitudes of the potentials is given for this mode
by 
\begin{equation*}
\mathbf{k}\cdot \mathbf{A}_{0}=\frac{\mu _{0}\varepsilon _{0}\omega }{1- 
\frac{\varepsilon _{0}}{\varepsilon }+i\frac{\sigma \left( 1+\gamma \right) 
}{\omega \varepsilon }}\left( \frac{1-\frac{\varepsilon _{0}}{\varepsilon }}{
1+\gamma ^{\prime }}+i\frac{\sigma }{\omega \varepsilon }\right) \phi _{0}.
\end{equation*}

Consequently, the amplitude of the scalar wave, $S_{0}=i\left( \mathbf{k}
\cdot \mathbf{A}_{0}-\mu _{0}\varepsilon _{0}\omega \phi _{0}\right) $, has
the expression{} 
\begin{equation}
S_{0}=i\frac{\mu _{0}\varepsilon _{0}\omega }{\left( 1+\gamma ^{\prime
}\right) }\left[ \frac{\gamma ^{\prime }\left( \frac{\varepsilon _{0}}{
\varepsilon }-1\right) -i\sigma \frac{\gamma }{\varepsilon \omega }\left(
1+\gamma ^{\prime }\right) }{1-\frac{\varepsilon _{0}}{\varepsilon }+i\frac{
\sigma \left( 1+\gamma \right) }{\omega \varepsilon }}\right] \phi _{0},
\label{S0phi0}
\end{equation}
and the corresponding longitudinal electric field amplitude satisfies 
\begin{equation}
\mathbf{k}\cdot \mathbf{E}_{0}=\frac{\omega \frac{\varepsilon _{0}}{
\varepsilon }}{1+i\frac{\sigma \left( 1+\gamma \right) }{\omega \varepsilon }
}S_{0}.  \label{kE0}
\end{equation}

For the case of a very good conductor, $\frac{\sigma }{\omega \varepsilon }
\gg 1$, relation (\ref{disprelS}) simplifies to 
\begin{equation*}
\frac{k^{2}}{\mu _{0}\varepsilon _{0}\omega ^{2}}=\frac{1}{1+\gamma },
\end{equation*}
and (\ref{S0phi0}) to 
\begin{equation*}
S_{0}=-i\frac{\gamma }{1+\gamma }\mu _{0}\varepsilon _{0}\omega \phi _{0}.
\end{equation*}

The corresponding longitudinal electric field amplitude is 
\begin{equation*}
E_{0}=-\frac{\gamma \varepsilon _{0}\omega ^{2}}{\sigma \left( 1+\gamma
\right) }\sqrt{\frac{\mu _{0}\varepsilon _{0}}{1+\gamma }}\phi _{0}.
\end{equation*}

These relations indicate that a small amplitude (for $\gamma \ll 1$) scalar
wave is possible in this case, propagating without decay in a good
conductor. Note that, consistently, the dissipated power density, $\sigma
E_{0}^{2}$, goes to zero for $\sigma \rightarrow \infty $.

For a dielectric without losses, $\sigma =0$, the dispersion relation is 
\begin{equation*}
\frac{k^{2}}{\mu _{0}\varepsilon _{0}\omega ^{2}}=\frac{1+\gamma ^{\prime } 
\frac{\varepsilon _{0}}{\varepsilon }}{1+\gamma ^{\prime }},
\end{equation*}
and the corresponding amplitudes of the scalar and of the longitudinal
electric field are 
\begin{eqnarray*}
S_{0} &=&-i\frac{\gamma ^{\prime }}{1+\gamma ^{\prime }}\mu _{0}\varepsilon
_{0}\omega \phi _{0}, \\
E_{0} &=&-i\gamma ^{\prime }\frac{\varepsilon _{0}}{\varepsilon }\omega 
\sqrt{\frac{\mu _{0}\varepsilon _{0}}{\left( 1+\gamma ^{\prime }\right)
\left( 1+\gamma ^{\prime }\frac{\varepsilon _{0}}{\varepsilon }\right) }}
\phi _{0}.
\end{eqnarray*}

As in the case of a conductor, a small amplitude (for $\gamma ^{\prime }\ll
1 $) scalar wave can propagate also in this case.

\section{Possible measurable effects}
\label{possible}

The general energy-momentum laws in AB electrodynamics were derived in
\cite{minotti2021quantum}. In relation to the interaction of fields and matter the following
expression for the power density on matter due to the fields was obtained 
\begin{equation}
w=\mathbf{j}\cdot \mathbf{E}-I\phi .  \label{wmediaAB}
\end{equation}
This is the microscopic expression that, apart from the usual term found in Maxwell electrodynamics, includes a term proportional to the extra source $I$. As a consequence, in media with non-conserved currents it would in principle be possible to observe the effect of this extra term.

\subsection{Detection of gauge-waves}
\label{detection}

As mentioned above, the possibility of material media with non-conserved
currents allows the detection of gauge-waves. This is so because according
to (\ref{wmediaAB}) the scalar potential of the wave can interact with any
extra source $I$ present in the medium. In the following a basic circuit
capable in principle of this detection is presented.

\begin{figure*}[ht]
\includegraphics[width=10cm]{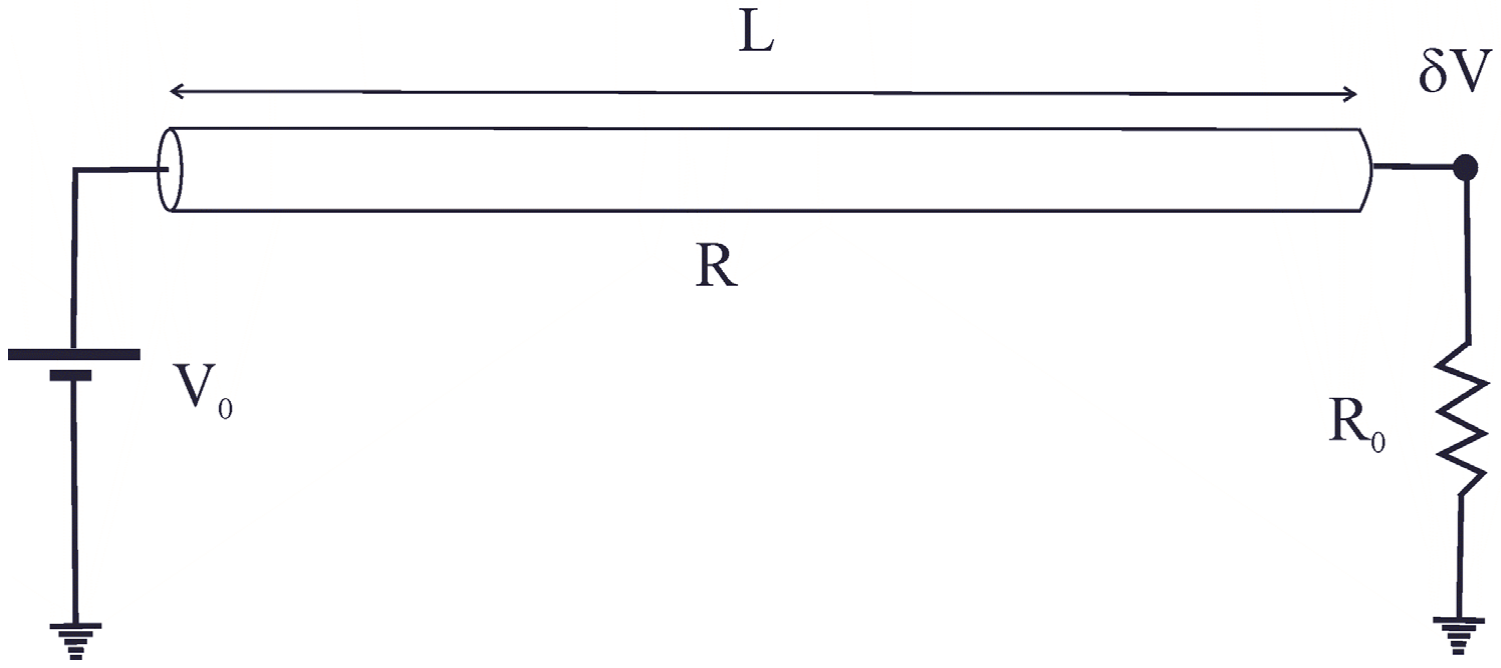}
\caption{Possible basic circuit for detection of gauge waves. The conductor of length 
$L$ and total electrical resistance $R$ sustains a continuous current $i_{0}=V_{0}/\left( R+R_{0}\right) $. The scalar potential of a gauge wave interacts with the not locally conserved part of the current and produces a varying voltage $\protect\delta V$ at the position denoted by the black dot.\label{qcircuit}}
\end{figure*}

In the conductor represented in Fig. (\ref{qcircuit}) there is a constant current of value $i_{0}=V_{0}/\left( R+R_{0}\right) $. The scalar potential $\phi $ of a gauge wave can then interact with the part of this current that is not locally conserved.

Since the gauge wave does not transport energy, and energy conservation is
already satisfied by the circulating permanent current, the law of
conservation of energy is expressed by (\ref{wmediaAB}) applied to the whole
circuit, expressed as 
\begin{equation*}
\int \left[ \delta \left( \mathbf{j}\cdot \mathbf{E}\right) -I\phi \right]
d^{3}x=0,
\end{equation*}
where the integral is extended to the whole volume of the conductor, and $
\delta \left( \mathbf{j}\cdot \mathbf{E}\right) =\delta \mathbf{j}\cdot 
\mathbf{E}+\mathbf{j}\cdot \delta \mathbf{E}$ represents the small change in the
corresponding magnitudes due to the interaction with the wave (it is assumed that $I\phi$, induced by the wave, is a small perturbation), while $\mathbf{E}$ and $\mathbf{j}$ are the values of the magnitudes when no gauge wave is present. In this way, denoting as $\delta i$ the change of current that arises due to interaction with the gauge wave, we can write
\begin{eqnarray*}
\int \delta \mathbf{j}\cdot \mathbf{E}d^{3}x &=&\delta i\left(
-V_{0}+Ri_{0}+R_{0}i_{0}\right) , \\
\int \mathbf{j}\cdot \delta \mathbf{E}d^{3}x &=&i_{0}\left( R\delta
i+R_{0}\delta i\right) ,
\end{eqnarray*}
so that
\begin{equation*}
\int \delta \left( \mathbf{j}\cdot \mathbf{E}\right) d^{3}x=\left[
-V_{0}+2\left( R+R_{0}\right) i_{0}\right] \delta i=V_{0}\delta i.
\end{equation*}

In order to evaluate $\int I\phi d^{3}x$ we use the general expression (\ref{FIXgeneral}) with $\phi $ in place of $F$, and extend the integration to
the whole circuit, so that 
\begin{equation*}
\int I\phi d^{3}x=\int \gamma \mathbf{j}\cdot \nabla \phi d^{3}x-\oint
\gamma \phi \mathbf{j}\cdot d\mathbf{S}\text{,}
\end{equation*}
where $\mathbf{j}$ is the current density in the circuit when there is no gauge wave. Note that the surface integral in this expression applies if the current that crosses the surface of the detecting element has a discontinuous component described by the same $\gamma $ coefficient as that
of the bulk of the detecting medium. In that case, after use of Gauss law we would obtain     
\begin{equation*}
\int I\phi d^{3}x=-\gamma \int \phi \nabla \cdot \mathbf{j}d^{3}x,
\end{equation*}
which is zero, since $\nabla \cdot \mathbf{j=0}$ for the DC circuit when no gauge wave is present. We must thus consider that the detecting medium is electrically fed by conductors with zero, or a much smaller value of $\gamma$ than that of the detecting element, so as to neglect the surface contribution, and finally obtain 
\begin{eqnarray*}
V_{0}\delta i &=&\int I\phi d^{3}x=\int \gamma \mathbf{j}\cdot \nabla \phi
d^{3}x \\
&=&\gamma i_{0}\int \nabla \phi \cdot \mathbf{n}dl=\gamma i_{0}\Delta \phi ,
\end{eqnarray*}
where the integral in the second line is preformed along the length of the
conductor, the unit vector $\mathbf{n}$ indicates the direction of the
circulating current, and the resulting $\Delta \phi $\ is the difference of
potential across the conductor length.

Using again $V_{0}=i_{0}\left( R+R_{0}\right) $ we thus have
\begin{equation*}
\delta i=\frac{\gamma }{\left( R+R_{0}\right) }\Delta \phi ,
\end{equation*}
or, from $\delta V=R_{0}\delta i$, 
\begin{equation*}
\delta V=\gamma \frac{R_{0}}{\left( R+R_{0}\right) }\Delta \phi .
\end{equation*}

\subsection{Antenna detector}
\label{antenna}

We consider as a relatively simple example a linear dipole of length $d$ and
negligible cross dimensions, fed by a source of voltage $V$\ of (angular)
frequency $\omega $. Taking the coordinate $z$ along the dipole, with the
origin at the dipole half point (where the power is fed to the antenna), a good approximation to the current circulating in the dipole is \cite{balanis2005}
\begin{equation*}
i\left( z,t\right) =i_{0}\sin \left( \frac{kd}{2}-k\left\vert z\right\vert
\right) \cos \left( \omega t\right) .
\end{equation*}

In terms of the antenna impedance, $Z_{A}=R_{A}+iX_{A}$, and that of the
source and circuit, $Z_{S}=R_{S}+iX_{S}$, the corresponding voltage is 
\begin{equation*}
V=i_{0}\left\vert Z_{A}+Z_{S}\right\vert \cos \left( \omega t+\varphi
\right) ,
\end{equation*} 
where 
\begin{equation*}
\tan \left( \varphi \right) =\left( X_{A}+X_{S}\right) /\left(
R_{A}+R_{S}\right) .
\end{equation*}

We consider that the condition for maximum power delivered to the antenna by
the circuit is fulfilled: $R_{S}=R_{A}$, $X_{S}=-X_{A}$, so that 
\begin{equation*}
V=2R_{A}i_{0}\cos \left( \omega t\right) .
\end{equation*}

If a gauge wave propagates along the $z$ direction with potential given by 
\begin{equation*}
\phi \left( z,t\right) =\phi _{0}\cos \left( k^{\prime }z-\omega ^{\prime
}t+\theta \right) ,
\end{equation*} 
the instantaneous power exchanged with the antenna is given by (the integral is extended to the antenna volume) 
\begin{eqnarray*}
-\int I\phi d^{3}x &=&\gamma \int \phi \nabla \cdot \mathbf{j}d^{3}x \\
&=&-\gamma ki_{0}\phi _{0}\cos \left( \omega t\right)
\int_{-d/2}^{d/2}sign\left( z\right) \cos \left( \frac{kd}{2}-k\left\vert
z\right\vert \right) \cos \left( k^{\prime }z-\omega ^{\prime }t+\theta
\right) dz \\
&=&2\gamma i_{0}\phi _{0}kk^{\prime }\frac{\cos \left( \frac{kd}{2}\right)
-\cos \left( \frac{k^{\prime }d}{2}\right) }{\left( k-k^{\prime }\right)
\left( k+k^{\prime }\right) }\cos \left( \omega t\right) \sin \left( \omega
^{\prime }t-\theta \right) .
\end{eqnarray*}

Conservation of energy requires that the additional $\delta i$ generated by
the interaction with the gauge wave satisfies 
\begin{equation*}
-V\delta i+\delta \left( 2R_{A}i^{2}\right) -\int I\phi d^{3}x=0,
\end{equation*} 
which, since $V=2R_{A}i$, reduces to 
\begin{equation*}
V\delta i-\int I\phi d^{3}x=0,
\end{equation*}
that determines the additional current in the circuit 
\begin{equation*}
\delta i=\frac{\gamma \phi _{0}}{R_{A}}kk^{\prime }\frac{\cos \left( \frac{kd 
}{2}\right) -\cos \left( \frac{k^{\prime }d}{2}\right) }{\left( k-k^{\prime
}\right) \left( k+k^{\prime }\right) }\sin \left( \omega ^{\prime }t-\theta
\right) .
\end{equation*}

The previous analysis is valid for $\omega ^{\prime }$ close to $\omega $ because the reactive component of the impedances was taken for simplicity as the same for both frequencies. For the case $\omega ^{\prime }=\omega $ the expression of the additional current simplifies to 
\begin{equation*}
\delta i=-\frac{\gamma \phi _{0}}{2R_{A}}\frac{kd}{2}\sin \left( \frac{kd}{2} 
\right) \sin \left( \omega t-\theta \right) .
\end{equation*}

It is interesting to note that a half-wavelength dipole, $d=\lambda /2$,
would be the best suited for the detection of a gauge wave at resonant
conditions.

\section{Radiating systems}
\label{radiating}

An important point for possible applications and/or experiments is the prediction of the fields radiated by a system in the context of the $I$-$\gamma$-$j$
model. We will consider the fields generated by harmonic in time currents in conductors in which polarization effects can be neglected.

We first recall that in AB electrodynamics the equations for the potentials derived from (\ref{ABeqsvaccum}) coincide with the corresponding Maxwell equations in the Lorentz gauge 
\begin{eqnarray*}
\frac{1}{c^{2}}\frac{\partial ^{2}\phi }{\partial t^{2}}-\nabla ^{2}\phi &=& 
\frac{\varrho }{\varepsilon _{0}}, \\
\frac{1}{c^{2}}\frac{\partial ^{2}\mathbf{A}}{\partial t^{2}}-\nabla ^{2} 
\mathbf{A} &=&\mu _{0}\mathbf{j},
\end{eqnarray*}
which after averaging result in equations with the same form for the macroscopic potentials, but with the sources replaced by the macroscopic ones, free plus molecular.

The general solutions in free space are given by 
\begin{eqnarray*}
\phi \left( \mathbf{x},t\right) &=&\frac{1}{4\pi \varepsilon _{0}}\int \frac{ 
\varrho \left( \mathbf{x}^{\prime },t^{\prime }\right) }{\left\vert \mathbf{x 
}^{\prime }-\mathbf{x}\right\vert }d^{3}x^{\prime }, \\
\mathbf{A}\left( \mathbf{x},t\right) &=&\frac{\mu _{0}}{4\pi }\int \frac{ 
\mathbf{j}\left( \mathbf{x}^{\prime },t^{\prime }\right) }{\left\vert 
\mathbf{x}^{\prime }-\mathbf{x}\right\vert }d^{3}x^{\prime },
\end{eqnarray*}
where $t^{\prime }=t-\left\vert \mathbf{x}^{\prime }-\mathbf{x}\right\vert
/c $. For harmonic in time fields and sources the corresponding Fourier-transformed-in-time amplitudes satisfy 
\begin{eqnarray*}
\phi \left( \mathbf{x},\omega \right) &=&\frac{1}{4\pi \varepsilon _{0}}
\int \frac{\varrho \left( \mathbf{x}^{\prime },\omega \right) }{R}\exp
\left( ikR\right) d^{3}x^{\prime }, \\
\mathbf{A}\left( \mathbf{x},\omega \right) &=&\frac{\mu _{0}}{4\pi }\int 
\frac{\mathbf{j}\left( \mathbf{x}^{\prime },\omega \right) }{R}\exp \left(
ikR\right) d^{3}x^{\prime },
\end{eqnarray*}
with $R=\left\vert \mathbf{x}^{\prime }-\mathbf{x}\right\vert $, and $ 
k=\omega /c$.

In the non-polarizable conductors considered the relation between the sources is given by (\ref{drhodtfinal}), which results for the Fourier amplitudes in 
\begin{equation*}
\varrho \left( \mathbf{x},\omega \right) =-i\frac{1+\gamma }{\omega }\nabla
\cdot \mathbf{j}.
\end{equation*}

We can thus write 
\begin{eqnarray*}
\phi \left( \mathbf{x},\omega \right) &=&-i\frac{1+\gamma }{4\pi \varepsilon
_{0}\omega }\int \frac{\nabla ^{\prime }\cdot \mathbf{j}}{R}\exp \left(
ikR\right) d^{3}x^{\prime } \\
&=&-i\frac{1+\gamma }{4\pi \varepsilon _{0}\omega }\oint \frac{\exp \left(
ikR\right) }{R}\mathbf{j}\cdot d\mathbf{S}^{\prime } \\
&&+i\frac{1+\gamma }{4\pi \varepsilon _{0}\omega }\int \mathbf{j}\left( 
\mathbf{x}^{\prime },\omega \right) \cdot \nabla ^{\prime }\left[ \frac{\exp
\left( ikR\right) }{R}\right] d^{3}x^{\prime }.
\end{eqnarray*}
Since the volume of integration contains all sources, no current traverses
its boundary, so that the surface integral is zero. Also, $\nabla ^{\prime
}R=-\nabla R$, allowing to finally obtain 
\begin{eqnarray*}
\phi \left( \mathbf{x},\omega \right) &=&-i\frac{1+\gamma }{4\pi \varepsilon
_{0}\omega }\nabla \cdot \int \frac{\mathbf{j}\left( \mathbf{x} ^{\prime
},\omega \right) }{R}\exp \left( ikR\right) d^{3}x^{\prime } \\
&=&-i\frac{\left(1+\gamma \right)c^{2}}{\omega }\nabla \cdot \mathbf{A}
\left( \mathbf{x},\omega \right) .
\end{eqnarray*}

In this way, all fields outside the sources can be determined from the solution for the vector potential alone 
\begin{equation}
\mathbf{A}\left( \mathbf{x},\omega \right) =\frac{\mu _{0}}{4\pi }\int \frac{
\mathbf{j}\left( \mathbf{x}^{\prime },\omega \right) }{R}\exp \left(
ikR\right) d^{3}x^{\prime },  \label{Axw}
\end{equation}
using the fields definitions in terms of the potentials, and the previous expression of $\phi \left( \mathbf{x},\omega \right) $, to obtain 
\begin{eqnarray*}
\mathbf{B}\left( \mathbf{x},\omega \right) &=&\nabla \times \mathbf{A}
\left( \mathbf{x},\omega \right) , \\
\mathbf{E}\left( \mathbf{x},\omega \right) &=&-\nabla \phi \left( \mathbf{x}
,\omega \right) +i\omega \mathbf{A}\left( \mathbf{x},\omega \right) \\
&=&i\frac{\left( 1+\gamma \right) c^{2}}{\omega }\nabla \times \mathbf{B}
\left( \mathbf{x},\omega \right) -i\gamma \omega \mathbf{A}\left( \mathbf{x}
,\omega \right) , \\
S\left( \mathbf{x},\omega \right) &=&-i\frac{\omega }{c^{2}}\phi \left( 
\mathbf{x},\omega \right) +\nabla \cdot \mathbf{A}\left( \mathbf{x},\omega
\right) \\
&=&-\gamma \nabla \cdot \mathbf{A}\left( \mathbf{x},\omega \right) .
\end{eqnarray*}

We can now use the usual approximations for the far field, or radiation region, in the solution of (\ref{Axw}), with sources located around the origin of coordinates, and $R\simeq \left\vert \mathbf{x}\right\vert - 
\mathbf{x}^{\prime }\cdot \mathbf{n}$, with $\mathbf{n}=\mathbf{x} 
/\left\vert \mathbf{x}\right\vert $. By further denoting $r\equiv \left\vert 
\mathbf{x}\right\vert $ the far-field term is 
\begin{equation}
\mathbf{A}\left( \mathbf{x},\omega \right) =\frac{\mu _{0}}{4\pi r}\exp
\left( ikr\right) \int \mathbf{j}\left( \mathbf{x}^{\prime },\omega \right)
\exp \left( -ik\mathbf{x}^{\prime }\cdot \mathbf{n}\right) d^{3}x^{\prime }.
\label{Afarfield}
\end{equation}

Since space derivatives of $r^{-1}$ and of $\mathbf{n}$ generate additional
factors $r^{-1}$, the only space derivatives of (\ref{Afarfield}) that
contribute in the radiation region are those of $\exp \left( ikr\right) $,
so that the general expression of the fields in the radiation region are 
\begin{subequations}
\label{BESphirad}
\begin{eqnarray}
\mathbf{B}\left( \mathbf{x},\omega \right)  &=&ik\mathbf{n}\times \mathbf{A}
\left( \mathbf{x},\omega \right) , \\
\mathbf{E}\left( \mathbf{x},\omega \right)  &=&-i\omega \left( 1+\gamma
\right) \mathbf{n}\times \left[ \mathbf{n}\times \mathbf{A}\left( \mathbf{x}
,\omega \right) \right] -i\gamma \omega \mathbf{A}\left( \mathbf{x},\omega
\right)   \notag \\
&=&i\omega \mathbf{A}_{\perp }\left( \mathbf{x},\omega \right) -i\gamma
\omega \mathbf{A}_{\parallel }\left( \mathbf{x},\omega \right) , \\
S\left( \mathbf{x},\omega \right)  &=&-ik\gamma \mathbf{n}\cdot \mathbf{A}
\left( \mathbf{x},\omega \right) , \\
\phi \left( \mathbf{x},\omega \right)  &=&\left( 1+\gamma \right) c\mathbf{n}
\cdot \mathbf{A}\left( \mathbf{x},\omega \right) 
\end{eqnarray}
\end{subequations} 
with $\mathbf{A}$ given by (\ref{Afarfield}), and where $\mathbf{A}_{\parallel }=\left( \mathbf{n}\cdot \mathbf{A}\right) 
\mathbf{n}$, $\mathbf{A}_{\perp }=\mathbf{A}-\mathbf{A}_{\parallel }$. Relation $ 
\omega =kc$ was also used. 

We thus see that the transverse components of the electromagnetic radiation
field generated by a given current distribution coincide with those
predicted by Maxwell electrodynamics. The additional terms are the
longitudinal component of the electric field and the scalar field.

The instantaneous energy flow has in AB electrodynamics the expression \cite{minotti2021quantum}
\begin{equation}
\mathbf{S}_{u}=\frac{1}{\mu _{0}}\left( \mathbf{E}\times \mathbf{B}-\phi
\nabla S+\mathbf{A}\frac{\partial S}{\partial t}\right) .
\label{energy_flux}
\end{equation}

In terms of the Fourier amplitudes the time-averaged energy flow is thus
expressed as (the asterisk indicates complex conjugation)
\begin{equation*}
\overline{\mathbf{S}}_{u}=\frac{1}{4\mu _{0}}\left[ \mathbf{E}^{\ast }\times 
\mathbf{B}+\mathbf{E}\times \mathbf{B}^{\ast }+ik\left( \phi S^{\ast }-\phi
^{\ast }S\right) +i\omega \left( \mathbf{A}S^{\ast }-\mathbf{A}^{\ast
}S\right) \right] .
\end{equation*}

Using expressions (\ref{BESphirad}) we readily obtain
\begin{equation*}
\overline{\mathbf{S}}_{u}=\frac{\omega ^{2}}{2\mu _{0}c}\left[ \mathbf{A}
_{\perp }\cdot \mathbf{A}_{\perp }^{\ast }-\gamma \left( 1+\gamma \right) 
\mathbf{A}_{\parallel }\cdot \mathbf{A}_{\parallel }^{\ast }\right] \mathbf{n
}.
\end{equation*}
The term that contains the perpendicular component $\mathbf{A}_{\perp }$
coincides with the usual Poynting vector in Maxwell electrodynamics. The interesting point is that the term containing the parallel component $\mathbf{A}_{\parallel }$ represents an "incoming" energy flow. This indicates that the tunneling currents lead to a "gain" of energy for the radiating system. 

As an example we take the finite-length dipole antenna previously
considered, for which 
\begin{equation*}
\mathbf{A}\left( \mathbf{x},\omega \right) =\mathbf{e}_{z}\frac{\mu _{0}i_{0} 
}{4\pi r}\exp \left( ikr\right) \int_{-d/2}^{d/2}\sin \left( \frac{kd}{2} 
-k\left\vert z^{\prime }\right\vert \right) \exp \left( -ikz^{\prime }\cos
\theta \right) dz^{\prime },
\end{equation*} 
where $\theta $ is the angle between $\mathbf{x}$ and the $z$ axis ($\mathbf{ 
e}_{z}$ is the unit vector along the $z$ axis). An elementary integration
gives 
\begin{equation*}
\mathbf{A}\left( \mathbf{x},\omega \right) =\frac{\mu _{0}i_{0}}{2\pi kr} 
\exp \left( ikr\right) \frac{\cos \left( \frac{kd}{2}\cos \theta \right)
-\cos \left( \frac{kd}{2}\right) }{\sin ^{2}\theta }\mathbf{e}_{z}.
\end{equation*}

The resulting magnetic far field has only azimuthal component 
\begin{equation*}
\mathbf{B}\left( \mathbf{x},\omega \right) =i\frac{\mu _{0}i_{0}}{2\pi r} 
\exp \left( ikr\right) \frac{\cos \left( \frac{kd}{2}\right) -\cos \left( 
\frac{kd}{2}\cos \theta \right) }{\sin \theta }\mathbf{e}_{\varphi },
\end{equation*}
which is transverse to the propagation direction $\mathbf{n}$.

The electric far field has the transverse $\theta $-component 
\begin{equation*}
E_{\theta }\left( \mathbf{x},\omega \right) =i\frac{i_{0}}{2\pi \varepsilon
_{0}cr}\exp \left( ikr\right) \frac{\cos \left( \frac{kd}{2}\right) -\cos
\left( \frac{kd}{2}\cos \theta \right) }{\sin \theta },
\end{equation*} 
and the radial, longitudinal component 
\begin{equation}
E_{r}\left( \mathbf{x},\omega \right) =i\gamma \frac{i_{0}}{2\pi \varepsilon
_{0}cr}\exp \left( ikr\right) \frac{\cos \left( \frac{kd}{2}\right) -\cos
\left( \frac{kd}{2}\cos \theta \right) }{\sin ^{2}\theta }\cos \theta .
\label{Eradlon}
\end{equation}

The scalar far field is 
\begin{equation}
S\left( \mathbf{x},\omega \right) =i\gamma \frac{\mu _{0}i_{0}}{2\pi r}\exp
\left( ikr\right) \frac{\cos \left( \frac{kd}{2}\right) -\cos \left( \frac{kd
}{2}\cos \theta \right) }{\sin ^{2}\theta }\cos \theta .  \label{Srad}
\end{equation}

The time-averaged energy flow is
\begin{equation*}
\overline{\mathbf{S}}_{u}=\frac{\mu _{0}i_{0}^{2}c}{8\pi ^{2}r^{2}}\left[ 
\frac{\cos \left( \frac{kd}{2}\cos \theta \right) -\cos \left( \frac{kd}{2}
\right) }{\sin ^{2}\theta }\right] ^{2}\left[ \sin ^{2}\theta -\gamma \left(
1+\gamma \right) \cos ^{2}\theta \right] \mathbf{n}.
\end{equation*}

\section{Conclusions}
\label{concl}

Building upon previous work, we have proven that the Aharonov-Bohm extended electrodynamics with the parameter choice $\lambda=1$ (corresponding to the so-called Feynman gauge Lagrangian in Maxwell theory) is the only theory of the electromagnetic field which allows a coupling to non-conserved currents and preserves the locality of field equations. In this theory the potentials $A^\mu$ are fixed, apart from a small class of residual gauge transformations, and can in principle be physically measured using anomalous conducting probes (i.e., probes made of materials in which the continuity $\partial_\mu j^\mu=0$ is violated at some points). As briefly recalled in Sect.\ \ref{phys}, such anomalies may exist, at the microscopic and quantum level, in various physical systems.

We can thus conclude that, although the exact local conservation of the current and the associated gauge symmetry are important and ``desirable'' properties, they do not constitute an absolute necessity, because the equations of the electromagnetic field make perfectly sense without them, and even become logically simpler.

On the other hand, we have proven that since it is still possible to introduce auxiliary, non-physical potentials with the familiar gauge-fixing choices, the extended theory is well compatible with Maxwell theory and inherits from it most of the usual computational techniques.

In order to demonstrate the consistency of the extended theory we have elaborated in detail on a phenomenological model of anomalous media, in which the continuity equation is replaced by the relation (derived from some microscopic fundamental hypotheses) $\partial_t \langle \rho \rangle + (1+\gamma) \nabla \cdot \langle \textbf{j} \rangle =0$. Here $\gamma$ is in general an adimensional parameter much smaller than 1; however, calculations of radiation emission in Sect.\ \ref{radiating} indicate that $\gamma$ can approach 1 from below without giving rise to any singularity. In Sect.\ \ref{wave-p} the propagation of plane waves in anomalous media has been computed.

Possible physical realizations of anomalous conductors are represented by molecular devices \cite{walz2015local,jensen2019current,garner2019helical,garner2020three}. We would like to point out that the parts of Ref.\ \cite{jensen2019current} most in line with our ideas are the Supplementary Materials, where the method by Wang et al.\ for the calculations of the non-local part of the current \cite{li2008definition} is applied. The authors find that a part of the current ``does not follow the bonds'' (i.e., the wavefunctions) also when the basis of states employed for the first-principles calculations is much enlarged. It should be stressed that first-principles  calculations give excellent results for observable quantities like energy spectra etc., so they appear to produce a good picture of the physics of the system, probably equivalent to that of an hypothetical renormalized quantum field theory, in which anomalies can play a role.

Another kind of macro-molecule exhibiting high conductivity and a charge transmission mechanism which is still not completely understood is graphene. Due to the chiral nature of quasi-particles in single and bi-layer graphene, quantum tunneling in these materials becomes highly anisotropic and markedly different from the case of normal, non-relativistic electrons. It is in fact a condensed-matter realization of the Klein tunneling, which raises a paradox concerning local charge conservation \cite{katsnelson2006chiral,allain2011klein}. The solutions of the paradox that have been proposed go beyond single-particle wavefunctions and involve multi-particle quantum field theory processes (see e.g.\ \cite{alkhateeb2021relativistic} and refs.). However, it is well known that chiral symmetry is often broken at the quantum level.

For this reason we believe that in the detection circuits for gauge waves proposed in Sect.\ \ref{possible} it might be convenient to use graphene or graphite as possible anomalous medium with non-zero extra-current $I$. However, a detailed discussion of those detectors concerning signal/noise sensitivity and possible error sources is beyond the scope of this paper and will be the subject of future work.

\bibliographystyle{ieeetr}
\bibliography{gwave}

\begin{thebibliography}{10}

\bibitem{bandos2020nonlinear}
I.~Bandos, K.~Lechner, D.~Sorokin, and P.~K. Townsend, ``{Nonlinear
  duality-invariant conformal extension of Maxwell’s equations},'' {\em
  Physical Review D}, vol.~102, no.~12, p.~121703, 2020.

\bibitem{sorokin2022introductory}
D.~P. Sorokin, ``Introductory notes on non-linear electrodynamics and its
  applications,'' {\em Fortschritte der Physik}, vol.~70, no.~7-8, p.~2200092,
  2022.

\bibitem{ohmura1956new}
T.~Ohmura, ``A new formulation on the electromagnetic field,'' {\em Progress of
  Theoretical Physics}, vol.~16, no.~6, pp.~684--685, 1956.

\bibitem{aharonov1963further}
Y.~Aharonov and D.~Bohm, ``Further discussion of the role of electromagnetic
  potentials in the quantum theory,'' {\em Physical Review}, vol.~130, no.~4,
  p.~1625, 1963.

\bibitem{van2001generalisation}
K.~Van~Vlaenderen and A.~Waser, ``Generalisation of classical electrodynamics
  to admit a scalar field and longitudinal waves,'' {\em Hadronic Journal},
  vol.~24, no.~5, pp.~609--628, 2001.

\bibitem{hively2012toward}
L.~Hively and G.~Giakos, ``Toward a more complete electrodynamic theory,'' {\em
  International Journal of Signal and Imaging Systems Engineering}, vol.~5,
  no.~1, pp.~3--10, 2012.

\bibitem{Modanese2017MPLB}
G.~Modanese, ``{Generalized Maxwell equations and charge conservation
  censorship},'' {\em Modern Physics Letters B}, vol.~31, p.~1750052, 2017.

\bibitem{modanese2017electromagnetic}
G.~Modanese, ``Electromagnetic coupling of strongly non-local quantum
  mechanics,'' {\em Physica B: Condensed Matter}, vol.~524, pp.~81--84, 2017.

\bibitem{arbab2017extended}
A.~Arbab, ``Extended electrodynamics and its consequences,'' {\em Modern
  Physics Letters B}, vol.~31, no.~09, p.~1750099, 2017.

\bibitem{hively2019classical}
L.~Hively and A.~Loebl, ``Classical and extended electrodynamics,'' {\em
  Physics Essays}, vol.~32, no.~1, pp.~112--126, 2019.

\bibitem{Hively_2021}
L.~M. Hively and M.~Land, ``Extended electrodynamics and {SHP} theory,'' {\em
  Journal of Physics: Conference Series}, vol.~1956, p.~012011, jul 2021.

\bibitem{minotti2021quantum}
F.~Minotti and G.~Modanese, ``Quantum uncertainty and energy flux in extended
  electrodynamics,'' {\em Quantum Reports}, vol.~3, no.~4, pp.~703--723, 2021.

\bibitem{minotti2022electromagnetic}
F.~Minotti and G.~Modanese, ``Electromagnetic signatures of possible charge
  anomalies in tunneling,'' {\em Quantum Reports}, vol.~4, no.~3, pp.~277--295,
  2022.

\bibitem{itzykson2012quantum}
C.~Itzykson and J.-B. Zuber, {\em Quantum field theory}.
\newblock Courier Corporation, 2012.

\bibitem{fradkin2021quantum}
E.~Fradkin, {\em Quantum field theory: an integrated approach}.
\newblock Princeton University Press, 2021.

\bibitem{woodside2009three}
D.~Woodside, ``{Three-vector and scalar field identities and uniqueness
  theorems in Euclidean and Minkowski spaces},'' {\em American Journal of
  Physics}, vol.~77, no.~5, pp.~438--446, 2009.

\bibitem{jimenez2011cosmological}
J.~Jim{\'e}nez and A.~Maroto, ``Cosmological magnetic fields from inflation in
  extended electromagnetism,'' {\em Physical Review D}, vol.~83, no.~2,
  p.~023514, 2011.

\bibitem{modanese2019design}
G.~Modanese, ``Design of a test for the electromagnetic coupling of non-local
  wavefunctions,'' {\em Results in Physics}, vol.~12, pp.~1056--1061, 2019.

\bibitem{minotti2021current}
F.~Minotti and G.~Modanese, ``Are current discontinuities in molecular devices
  experimentally observable?,'' {\em Symmetry}, vol.~13, no.~4, p.~691, 2021.

\bibitem{minotti2023aharonov}
F.~Minotti and G.~Modanese, ``{Aharonov--Bohm Electrodynamics in Material
  Media: A Scalar e.m. Field Cannot Cause Dissipation in a Medium},'' {\em
  Symmetry}, vol.~15, no.~5, p.~1119, 2023.

\bibitem{morawetz2019chiral}
K.~Morawetz, ``{Chiral anomaly in Weyl systems: No violation of classical
  conservation laws},'' {\em Physics Letters A}, vol.~383, no.~12,
  pp.~1362--1363, 2019.

\bibitem{morawetz2019weyl}
K.~Morawetz, ``{Weyl systems: anomalous transport normally explained},'' {\em
  The European Physical Journal B}, vol.~92, no.~8, pp.~1--15, 2019.

\bibitem{cheng1984gauge}
T.-P. Cheng and L.-F. Li, {\em Gauge theory of elementary particle physics}.
\newblock Clarendon Press Oxford, 1984.

\bibitem{parameswaran2014probing}
S.~Parameswaran, T.~Grover, D.~Abanin, D.~Pesin, and A.~Vishwanath, ``Probing
  the chiral anomaly with nonlocal transport in three-dimensional topological
  semimetals,'' {\em Physical Review X}, vol.~4, no.~3, p.~031035, 2014.

\bibitem{walz2015local}
M.~Walz, A.~Bagrets, and F.~Evers, ``Local current density calculations for
  molecular films from ab initio,'' {\em Journal of Chemical Theory and
  Computation}, vol.~11, no.~11, pp.~5161--5176, 2015.

\bibitem{jensen2019current}
A.~Jensen, M.~Garner, and G.~Solomon, ``When current does not follow bonds:
  Current density in saturated molecules,'' {\em The Journal of Physical
  Chemistry C}, vol.~123, no.~19, pp.~12042--12051, 2019.

\bibitem{garner2019helical}
M.~Garner, A.~Jensen, L.~Hyllested, and G.~Solomon, ``Helical orbitals and
  circular currents in linear carbon wires,'' {\em Chemical science}, vol.~10,
  no.~17, pp.~4598--4608, 2019.

\bibitem{garner2020three}
M.~Garner, W.~Bro-J{\o}rgensen, and G.~Solomon, ``Three distinct torsion
  profiles of electronic transmission through linear carbon wires,'' {\em The
  Journal of Physical Chemistry C}, vol.~124, no.~35, pp.~18968--18982, 2020.

\bibitem{lenzi2008solutions}
E.~Lenzi, B.~de~Oliveira, L.~da~Silva, and L.~Evangelista, ``{Solutions for a
  Schr{\"o}dinger equation with a nonlocal term},'' {\em Journal of
  Mathematical Physics}, vol.~49, no.~3, p.~032108, 2008.

\bibitem{lenzi2008fractional}
E.~Lenzi, B.~De~Oliveira, N.~Astrath, L.~Malacarne, R.~Mendes, M.~Baesso, and
  L.~Evangelista, ``Fractional approach, quantum statistics, and
  non-crystalline solids at very low temperatures,'' {\em The European Physical
  Journal B-Condensed Matter and Complex Systems}, vol.~62, no.~2,
  pp.~155--158, 2008.

\bibitem{latora1999superdiffusion}
V.~Latora, A.~Rapisarda, and S.~Ruffo, ``Superdiffusion and out-of-equilibrium
  chaotic dynamics with many degrees of freedoms,'' {\em Physical Review
  Letters}, vol.~83, no.~11, p.~2104, 1999.

\bibitem{caspi2000enhanced}
A.~Caspi, R.~Granek, and M.~Elbaum, ``Enhanced diffusion in active
  intracellular transport,'' {\em Physical Review Letters}, vol.~85, no.~26,
  p.~5655, 2000.

\bibitem{chamon1997nonlocal}
L.~Chamon, D.~Pereira, M.~Hussein, M.~Ribeiro, and D.~Galetti, ``Nonlocal
  description of the nucleus-nucleus interaction,'' {\em Physical Review
  Letters}, vol.~79, no.~26, p.~5218, 1997.

\bibitem{balantekin1998green}
A.~Balantekin, J.~Beacom, {\em et~al.}, ``Green's function for nonlocal
  potentials,'' {\em Journal of Physics G: Nuclear and Particle Physics},
  vol.~24, no.~11, p.~2087, 1998.

\bibitem{laskin2002fractional}
N.~Laskin, ``{Fractional Schr{\"o}dinger equation},'' {\em Physical Review E},
  vol.~66, no.~5, p.~056108, 2002.

\bibitem{wei2016comment}
Y.~Wei, ``{Comment on ``Fractional quantum mechanics'' and ``Fractional
  Schr{\"o}dinger equation''},'' {\em Physical Review E}, vol.~93, no.~6,
  p.~066103, 2016.

\bibitem{modanese2018time}
G.~Modanese, ``Time in quantum mechanics and the local non-conservation of the
  probability current,'' {\em Mathematics}, vol.~6, no.~9, p.~155, 2018.

\bibitem{Minotti-PRE-2000}
F.~O. Minotti, ``Self-consistent derivation of subgrid stresses for large-scale
  fluid equations,'' {\em Phys. Rev. E}, vol.~61, pp.~429--434, Jan 2000.

\bibitem{balanis2005}
C.~A. Balanis, {\em Antenna Theory: Analysis and Design}.
\newblock Wiley-Interscience, 2005.

\bibitem{li2008definition}
C.~Li, L.~Wan, Y.~Wei, and J.~Wang, ``Definition of current density in the
  presence of a non-local potential,'' {\em Nanotechnology}, vol.~19, no.~15,
  p.~155401, 2008.

\bibitem{katsnelson2006chiral}
M.~I. Katsnelson, K.~S. Novoselov, and A.~K. Geim, ``Chiral tunnelling and the
  klein paradox in graphene,'' {\em Nature physics}, vol.~2, no.~9,
  pp.~620--625, 2006.

\bibitem{allain2011klein}
P.~E. Allain and J.-N. Fuchs, ``Klein tunneling in graphene: optics with
  massless electrons,'' {\em The European Physical Journal B}, vol.~83,
  pp.~301--317, 2011.

\bibitem{alkhateeb2021relativistic}
M.~Alkhateeb, X.~G. de~La~Cal, M.~Pons, D.~Sokolovski, and A.~Matzkin,
  ``Relativistic time-dependent quantum dynamics across supercritical barriers
  for klein-gordon and dirac particles,'' {\em Physical Review A}, vol.~103,
  no.~4, p.~042203, 2021.

\end{thebibliography}

\end{document}